\documentclass[referee,useAMS,usenatbib]{mn2e}
\usepackage{graphicx,graphics,amsmath,amssymb,mathrsfs}
\usepackage{subeqn, wasysym}
\usepackage[T1]{fontenc}
\usepackage{aecompl} 

\newcommand{\nin}{\noindent}
\newcommand{\beq}{\begin{equation}}
\newcommand{\eeq}{\end{equation}}

\newcommand{\bfa}{\mbox{\boldmath $a$}}
\newcommand{\bfb}{\mbox{\boldmath $b$}}
\newcommand{\bfr}{\mbox{\boldmath $r$}}
\newcommand{\bfu}{\mbox{\boldmath $u$}}
\newcommand{\bfA}{\mbox{\boldmath $A$}}

\newcommand{\bfX}{\mbox{\boldmath $X$}}
\newcommand{\bfx}{\mbox{\boldmath $x$}}
\newcommand{\bfv}{\mbox{\boldmath $v$}}

\newcommand{\cendot}{\mbox{\boldmath $\cdot$}}
\newcommand{\p}{\mbox{$\partial$}}
\newcommand{\rmd}{\mbox{$\rm d$}}
\newcommand{\tk}{\mbox{$T_{\rm kep}$}}
\newcommand{\ts}{\mbox{$T_{\rm sec}$}}
\newcommand{\scrw}{{\cal W}}
\newcommand{\scrc}{{\cal C}}

\newcommand{\scrd}{{\cal D}}

\newcommand{\scrr}{{\cal R}}

\begin{document}

\title[Stellar Dynamics around a Massive Black Hole~II]{Stellar Dynamics around a Massive Black Hole~II:\\Resonant Relaxation}

\author[Sridhar \& Touma]{S.~Sridhar$^{1,3}$ and Jihad~R.~Touma$^{2,4}$\\
$^{1}$ Raman Research Institute, Sadashivanagar, Bangalore 560 080, India\\
$^{2}$ Department of Physics, American University of Beirut, PO Box 11-0236, Riad El-Solh, Beirut 11097 2020, Lebanon\\
$^{3}$ ssridhar@rri.res.in $^{4}$ jt00@aub.edu.lb\\
} 
\maketitle
\vspace{-2em}
\begin{abstract}
We present a first--principles theory of Resonant Relaxation (RR) of a low mass stellar system orbiting a more massive black hole (MBH). We first extend the kinetic theory of \citet{gil68} to include the Keplerian field of a black hole of mass $M_\bullet$. Specializing to a Keplerian stellar system of mass $M \ll M_\bullet$, we use the orbit--averaging method of 
\citet{st15} to derive a kinetic equation for RR. This describes the 
collisional evolution of a system of $N \gg 1$ Gaussian Rings in a reduced 5--dim space, under the combined actions of self--gravity, 1~PN and 1.5~PN 
relativistic effects of the MBH and an arbitrary external potential. 
In general geometries RR is driven by both apsidal and nodal resonances, so the distinction between scalar--RR and vector--RR disappears. The system passes through a sequence of quasi--steady secular collisionless equilibria, driven by irreversible 2--Ring correlations that accrue through gravitational interactions, both direct and collective. This correlation function is related to a `wake function', which is the linear response of the system to the perturbation of a chosen Ring.  The wake function is easier to appreciate, and satisfies a simpler equation, than the correlation function. We discuss general implications for the interplay of secular dynamics and non--equilibrium statistical mechanics in the evolution of Keplerian stellar systems toward secular thermodynamic equilibria, and set the stage for applications to the RR of axisymmetric discs in Paper~III. 
\end{abstract}


\section{Introduction}

Star clusters around massive black holes (MBH) in galactic nuclei are dense stellar systems in which stellar orbital motions are fast enough 
for relaxation processes to have significantly modified their structure.
Relaxation is driven by gravitational ``collisions'' between stars, leading to exchanges of orbital energies and angular momenta. Classical two--body relaxation results in the diffusion of both energy and angular momentum, over the two-body relaxation time scale $T_{\rm relax}$ 
\citep{cha42, cha43a, cha43b, bt08}. In its fully developed form two--body relaxation can feed the MBH with stars, and lead to the formation of a stellar density cusp around the MBH \citep{bw76,ck78}. But for the stellar systems observed in galactic nuclei, $T_{\rm relax}$ can exceed the Hubble time; then two--body relaxation would have had negligible effect. \citet{rt96} --- hereafter RT96 --- proposed a more efficient mechanism, \emph{Resonant Relaxation} (RR), that could be in operation in stellar systems with globally degenerate (or resonant) orbital frequencies. RR 
does not enhance the energy relaxation rate but promotes more efficient relaxation of orbital angular momentum distribution. When this happens 
in a low mass stellar system orbiting a MBH, stars whose orbits have become  sufficiently eccentric can fall into the MBH, feeding it mass, 
energy and angular momentum. RR is particularly important in a low--mass stellar system around a MBH, and is a result of the degeneracy of the Kepler problem. As the radial and azimuthal periods of a Keplerian ellipse are equal, orbital precession is suppressed; only the smaller non--Keplerian forces, such as cluster self--gravity, general relativistic effects and 
external gravitational sources contribute to orbital precession. This gives rise to a pattern of interactions between Keplerian orbits which persists over the longer time scales of precession.  It is this extended period of interplay, or coherence, between Keplerian orbits that makes RR more efficient than classical two--body relaxation. 

\medskip
\noindent
{\bf \emph{Sketch of the Rauch--Tremaine model}:} 
Let us consider a stellar system of size $R$ and total mass $M$, consisting of $N \gg 1$ stars (each of mass $m_\star = M/N$), orbiting a MBH of mass $M_\bullet\, $. Stellar orbits are determined by the combined gravitational effects of the Keplerian potential of the MBH and cluster self--gravity, general relativistic effects and any external gravitational sources. The stellar system will be referred to as Keplerian \citep{st15} when the non--Keplerian forces are much smaller than the Keplerian field of the MBH. Therefore, for an isolated Keplerian stellar system the mass ratio must be small, $\varepsilon \equiv M/M_\bullet \ll 1\,$, and its size must be large enough, $R \gg r_\bullet\,$, where $r_\bullet \equiv 2GM_\bullet/c^2$ is the Schwarzschild radius of the MBH.  In 
two--body relaxation random gravitational encounters between stars, with a coherence time of order the short Kepler orbital timescale $\tk = 2\pi\left(R^3/GM_\bullet\right)^{1/2}\,$, build up to order unity over times $T_{\rm relax} = N\varepsilon^{-2} T_{\rm kep}\,$.\footnote{Logarithmic corrections to $T_{\rm relax}$ have been neglected.} As noted earlier $T_{\rm relax}$ can exceed the Hubble time, whereas RR works more efficiently because of the longer coherence times which we now estimate. Following RT96 we consider RR in a non--relativistic Keplerian stellar system, for which $(r_\bullet/R) \ll \varepsilon \ll 1\,$. Then orbits evolve slowly over secular timescales $\ts = \varepsilon^{-1} \tk\,$, which is longer than Kepler orbital times. Over several $\ts$, stellar orbits can be thought of as \emph{Gaussian Rings} (see Paper~I), which are Keplerian ellipses of fixed semi--major axes that can deform in the mean self--gravitational field of the cluster. A secular collisionless theory of this system was worked out in Paper~I, and applies in the limit $N \to\infty$ while $M = Nm_\star$ is held constant. For realistic systems $N$ is large but finite, say $N \sim 10^5 - 10^7$, and it is this granularity that drives RR.

As the semi--major axes of Gaussian Rings do not change, there is no exchange of Keplerian energies during secular gravitational encounters.\footnote{The Keplerian orbital energies diffuse over the longer time scale $T_{\rm relax}$.} But mutual torquing can exchange angular momentum between Rings leading to diffusion of angular momenta. The diffusion of a vectorial quantity like angular momentum can be complicated, and depends on the geometry and orbital structure of the stellar system.  According to RT06 the physics of this process can be understood by considering two extreme cases: \emph{scalar--RR} which is the diffusion of the magnitude of angular momenta (and hence the eccentricity of stellar orbits); and \emph{vector--RR} which is the diffusion of the direction of angular momenta. Diffusion of the magnitude of angular momenta occurs as a result of the mutual torquing of Gaussian Rings whose apsides precess over times $\ts$. Hence the coherence time for these secular encounters is $\sim \ts$, which is much longer than $\tk$. The magnitude of the angular momentum can be thought of as varying linearly in time for times shorter than $\ts$, and random--walking over longer times. The timescale for scalar--RR turns out to be  $T_{\rm res} = N\ts = \varepsilon T_{\rm relax}\,$, which is shorter than $T_{\rm relax}$. Vector--RR can occur in near--spherical systems, even in the absence of a MBH, when the apsides of stellar orbits precess fast (say, when the stellar system is immersed in an external density cusp) and apsidal resonances have negligible effect. Then orbital eccentricities are conserved, in addition to the semi--major axes. Averaging over the fast apsidal precession, each stellar orbit can be replaced by an axisymmetric annulus whose inner/outer radii are equal to the (constant) peri/apo centre distances. Diffusion of the orientation of orbits occurs as a result of the mutual torquing of the stellar annuli with  coherence times, $\sim N^{1/2}\ts$. In a Keplerian stellar system it turns out that the vector--RR timescale is also of order the coherence time, giving $T^{\rm v}_{\rm res} = N^{1/2}\ts = N^{-1/2}T_{\rm res}$, which is shorter than even the scalar--RR timescale. Thus there emerges the notion of a hierarchy of relaxation processes, with longer coherence times corresponding to shorter relaxation times: (a) the Keplerian orbital timescale and classical two--body relaxation; (b) the slow secular apse precession timescale and scalar--RR; (c) the slower secular nodal precession timescale and vector--RR. For general stellar distributions the distinction between scalar--RR and vector--RR can get blurred.

\medskip
\noindent
{\bf \emph{Resonant Relaxation over the years}:} 
RR has been explored extensively through both numerical simulations and stochastic modeling, a trend that was also initiated by RT96: restricted 
$N$--Body simulations (splitting stars into field and test stars) and $N$--wire simulations were used to demonstrate that RR is more efficient than two--body relaxation; the efficiency was characterized in both coherent/linear and diffusive regimes, by fitting for parameters that were left free in their random walk model. \citet{ha06} revisited the problem of the feeding of a MBH with RR acting as the driving mechanism, instead of the classical two--body relaxation considered by \citet{bw76}. Using the RT96 model they parametrized the strength of RR in a Fokker--Planck equation (which is averaged over angular momentum) for the distribution of energies, with loss cone boundary conditions. They could constrain the fluxes into the MBH, the energies of stars that contribute to it, and enhancement in gravitational wave events. They also studied the implications of the process for the Galactic Centre and concluded that the kinematics of stars was consistent with the workings of RR. \citet{gh07} used the wire--approximation of RT96 and random sampling to explore the RR time scale as a function of eccentricity, and concluded that RR is more efficient on eccentric orbits. \citet{kea10} used small--scale Newtonian $N$--body simulations to determine the efficiency of RR with regard to the inspiral of compact remnants into the MBH and emission of gravitational waves, and concluded that RR will increase the rates of inspiral events by a factor of a few over two-body relaxation.

\citet{mhl11} generalized \citet{ha06} in two significant ways by using: (a) An autoregressive moving average model for RR (instead of using the RT06 random walk model) that is constrained by restricted $N$--body simulations; (b) A Fokker--Planck equation that kept track of both energy and angular momentum. Their solution allowed them to account for a density core in the Galactic Center, and to conclude that binary disruption makes for resonantly relaxed orbits which are too eccentric for the observed S--stars. \citet{kt11} explored an analytical model of vector--RR to investigate possible warps in the distribution of stars at the Galactic Centre. Vector--RR has also been studied by \citet{kt15} through numerical simulations of annuli--annuli interactions, and a random--walk model of orbit normals on a sphere. They found that the dominant torques were between stars with radially overlapping orbits, and that the vector--RR rate increased rapidly for highly eccentric ($e\gtrsim 0.8$) orbits. \citet{tt14} determined maximum entropy equilibria of self--gravitating Keplerian discs, for the case where all the particles have equal semi--major axes. Their numerical investigations revealed broken--symmetry states which could be lopsided and uniformly precessing. 

Using extensive post--Newtonian $N$--body simulations, \citet{mamw11}, argued that general relativistic precession offers a severe barrier to the capture of stars by the MBH, thereby quenching RR. They considered dynamical mechanisms for barrier penetration and assessed the rate of gravitational wave events they enable. This is the most serious attempt at studying the workings of RR through $N$-body simulations, and stands as a benchmark for confrontation with any fundamental theory of RR in the presence of general relativistic corrections. \citet{ba14,ba15} have also studied the problem of the quenching of RR torques by general relativistic precession; they model the background potential as a correlated Gaussian noise, and study the 
stochastic process through a Fokker--Planck equation and Monte Carlo simulations. \citet{hpm14} made extensive post--Newtonian, restricted $N$--body simulations with a view to understanding the behavior of S--stars in the Galactic Centre. They identified three regimes of relaxation: non--resonant, resonant and anomalous, and recovered functional forms for the diffusion coefficients in all three regimes. These results were used in a Fokker--Planck equation to obtain the steady-state distribution of angular momentum for orbits near the MBH. \citet{m15a,m15b,m15c} developed a numerical algorithm of the Cohn--Kulsrud type for solving a Fokker-Planck equation in both energy and angular momentum, which used diffusion coefficients describing the effects of two--body relaxation, RR, anomalous relaxation and energy loss due to emission of gravitational waves; loss of stars to the MBH was also included. The algorithm was applied to recover time--evolving distribution functions in galactic nuclei.

\medskip
\noindent
{\bf \emph{Approach to a Kinetic Theory of Resonant Relaxation}:} The brief overview of developments in RR given above is, admittedly, far from exhaustive; rather it is meant to be illustrative of what we perceive as some major trends over the past two decades. Numerical simulations have grown in sophistication; in quality and size, as well as in their ability to include physical effects such as general relativistic precession and emission of gravitational waves. In contrast theoretical understanding of 
RR is based on various parametrized stochastic models whose parameters are recovered from the numerical simulations. It is necessary to develop 
a new framework for RR, one whose foundations are based on rigorous theories 
of non--equilibrium statistical mechanics. We begin by recalling the 
theoretical basis underlying the standard treatment of classical two--body 
relaxation.

Two--body relaxation has been studied extensively both analytically and numerically. The historical beginnings of the kinetic theory of stellar systems are rooted in the problem of the evolution of globular clusters. \citet{cha42,cha43a,cha43b} derived a Fokker--Planck equation by considering gravitational encounters in an infinite homogeneous system of stars. The derivation of the diffusion and friction coefficients was simplified by 
\citet{csr50, gnr56, rmj57}. The general kinetic equation for inhomogeneous stellar systems was derived by \citet{gil68}; this is rigorous and based on a systematic $1/N$ expansion of the BBGKY (Bogoliubov, Born, Green, Kirkwood, Yvon) equations of non--equilibrium statistical mechanics. \citet{hey10} revisited the problem using action--angle variables (assuming an integrable system), and proved that the kinetic equation satisfies an H--theorem. The most difficult parts of the theory are certain polarization terms arising from collective interactions. A first simplification of the kinetic equation was achieved by \citet{ps82} through the neglect of the polarization terms --- see also \citet{cha12,cha13}.\footnote{This simplification is analogous to the passage from the equations of \citet{bal60} and 
\citet{len60} to the \citet{l36} equations, for electrostatic plasmas.} There has been some work on the applications of the kinetic equation (in the Gilbert form, or the reduced Polyachenko--Shukhman form), but we do not review this literature. We note, however, that the kinetic equation applies to inhomogeneous systems for which computing interactions between realistic stellar orbits can be complicated. One can pass from \citet{ps82} to the more tractable and well--investigated Fokker--Planck description of classical two--body relaxation. This involves the step taken originally by Chandrasekhar, in which gravitational encounters are estimated by assuming that stars move with constant velocities. Thus we are comforted that the journeyman theory of two--body relaxation, which is textbook material \citep{bt08}, rests on the firm foundations of \citet{gil68} through descent via \citet{ps82}.

But we have seen that two--body relaxation is not efficient in galactic nuclei with MBHs, and we must consider RR. To the best of our knowledge, 
the stochastic models used to interpret numerical simulations of RR are 
all \emph{ad hoc}. Our goal is to develop a first--principles theory of RR, that is based on \citet{gil68}. For simplicity we consider a stellar 
system consisting of equal mass stars; the extension to a stellar system 
with a range of masses is straightforward, but this will not be treated here. Our focus is on presenting the derivation of the kinetic equation 
governing RR in detail, so all the developments in this paper are of a formal nature. Applications of our theory will be presented in separate papers. We begin in \S~2 with an outline of Gilbert's theory of the collisional relaxation of general stellar systems, adding to it arbitrary external sources of gravity. The theory is based on a $1/N$ expansion: at order unity we get the collisionless Boltzmann equation (CBE), and collision terms emerge at order $1/N\,$. We cast the equations in Poisson Bracket form by defining various potentials. In \S~3 the external potential is taken to be that of a MBH, and a canonical transformation is effected to coordinates centered on the MBH. Another canonical transformation expresses the Gilbert equations in the Delaunay variables, which are the natural action--angle variables for the Kepler problem. No approximations have been made and the equations apply to a stellar system around a MBH for any mass ratio $\varepsilon$. In \S~4 we consider the application to Keplerian stellar systems for which $\varepsilon \ll 1$. We begin with the collisionless theory and recall the results of Paper~I on the secular limit, obtained using the method of multiple scales. Then we move to the  collisional theory and orbit--average the Gilbert equations; the calculations are straightforward, with details given in the Appendix. The result is a kinetic equation for RR as envisioned by RT96. In \S~5 we supplement the mean--self gravity of the system by secular relativistic corrections (up to 1.5 post--Newtonian order) and external sources of gravity. Thus we arrive at a general kinetic equation for RR. Some general physical features of this equation are discussed in \S~6, and concluding remarks are offered in \S~7.

\section{General setting for collisional relaxation}

\subsection{Outline of Gilbert's theory}

We begin with a brief account of the theory of collisional relaxation of \citet{gil68}. The stellar system (or cluster) consists of a large number 
$N$ of stars, each of mass $m_\star$, and total mass $M=Nm_\star$. Let $\bfx_i$ and $\bfv_i$ be the position and velocity of the $i$th star (for $i=1,\ldots,N$), with respect to an inertial frame. The phase space of the system is $6N$--dim with coordinates $\Gamma_i \equiv (\bfx_i, \bfv_i)$. We also use $\rmd\Gamma_i = \rmd^3\bfx_i\rmd^3\bfv_i$ and write the $6N$--dim volume element as $\rmd\Gamma_1\ldots\rmd\Gamma_N$. We recall that, in the standard statistical formulation \citep{lp81}, the system is described by an $N$--particle  distribution function (DF) $f^{(N)}(\Gamma_1,\ldots, \Gamma_N, t)$ which is a symmetric function of $\Gamma_1,\ldots,\Gamma_N$. Reduced $s$--particle DFs $f^{(s)}(\Gamma_1,\ldots, \Gamma_s, t)$ are derived for $s=1,\ldots,(N-1)$, by integrating $f^{(N)}$ over the coordinates $\Gamma_{s+1},\ldots,\Gamma_N$. The $N$--particle DF is a probability distribution in the $6N$--dim phase space $\{\Gamma_1,\ldots,\Gamma_N\}$; hence all the $s$--particle DFs are probability distributions in their respective $6s$--dim phase spaces $\{\Gamma_1,\ldots,\Gamma_s\}$. Evolution equations for the $f^{(s)}$ can be obtained, by integrating the Liouville equation over these very coordinates. This is the BBGKY hierarchy of equations, in which the equation for $f^{(1)}$ depends on $f^{(2)}$, and so on up to $f^{(N)}$. For a large $N$ system the chain of equations is very long, and is usually truncated through ordering in a physically relevant small parameter. 

Drawing on earlier work in plasma physics \citep{bal60,len60}, Gilbert achieved closure of the BBGKY equations for stellar systems through 
power--series expansion in the small parameter $(1/N)$. With notation $\,\Gamma = (\bfx, \bfv)$, $\,\Gamma' = (\bfx', \bfv')$ and $\,\Gamma'' = (\bfx'', \bfv'')$, the 2--particle DF can be expanded as: $\,f^{(2)}(\Gamma,\Gamma', t)=  f^{(1)}(\Gamma, t)f^{(1)}(\Gamma', t) \,+\, (1/N)f^{(2)}_{\rm irr}(\Gamma, \Gamma', t) \,+\, \ldots\,$, where $\,f^{(2)}_{\rm irr}\,$ is the irreducible part of the 2--particle correlation, satisfying $\int f^{(2)}_{\rm irr}\,\rmd\Gamma \,=\, \int f^{(2)}_{\rm irr}\,\rmd\Gamma' \,=\, 0$. It is straightforward to verify that $f^{(3)}$ is determined uniquely in terms of $f^{(1)}$ and $f^{(2)}_{\rm irr}$, to first order in $1/N$. Hence the BBGKY hierarchy closes at this order, and provides two coupled equations for $f^{(1)}$ and $f^{(2)}_{\rm irr}$ which describe the collisional evolution of a self--gravitating system. Irreversibility arises through the ``adiabatic turn--on'' initial conditions, for which $f^{(2)}_{\rm irr}$ vanishes in the distant past. Henceforth we drop the subscript ``1'', and write the 1--particle DF as $f(\Gamma, t)$. The acceleration of a test particle at $(\bfx, t)$ by the mean--field is: 
\beq
\bfa(\bfx, t)\;=\; \int f(\Gamma', t)\,\bfb(\bfx, \bfx')\,\rmd\Gamma'\,,\qquad\mbox{where}\qquad\bfb(\bfx, \bfx')
\;=\; GM\frac{\bfx'-\bfx}{\,\left|\bfx'-\bfx\right|^3\,} 
\label{meanaccn}
\eeq
is the (scaled) Newtonian ``bare'' inter--particle acceleration. External sources of gravity (a MBH, an in--falling gas cloud or star cluster) can also be included in the formalism, although these are absent in \citet{gil68}. Let $\bfa^{\rm e}(\bfx, t)$ be the acceleration of the test star
due to external sources. Then the equation for $f(\bfx, \bfv, t)$ can be written as:
\beq
\frac{\p f}{\p t} \;+\; \bfv \cendot \frac{\p f}{\p \bfx}
\;+\; \left(\bfa + \bfa^{\rm e}\right)\cendot\frac{\p f}{\p \bfv} 
\;=\; \frac{1}{N}\bfa\cendot\frac{\p f}{\p \bfv} \;-\; 
\frac{1}{N}\int \bfb(\bfx, \bfx')\cendot\frac{\p f^{(2)}_{\rm irr}(\Gamma, \Gamma', t)}{\p \bfv}\,\rmd\Gamma'\,.
\label{gil-org}
\eeq 
Mean--field theory (MFT) corresponds to the limiting case when $N\to\infty$, $m_\star \to 0$ with $Nm_\star = M = \mbox{mass of the system which is held constant}$. The accelerations $\bfb(\bfx, \bfx')$, $\,\bfa(\bfx, t)$ and $\bfa^{\rm e}(\bfx, t)$ are all well--defined in this limit. However, both terms on the right hand side vanish in the limit. Hence the MFT of a stellar system is completely described by the 1--particle DF, $f(\bfx, \bfv, t)$, whose time evolution is governed by the \emph{collisionless Boltzmann equation} (CBE). When $N$ is large but not infinite, the accrual of small changes induced in $f$, by the right hand side of eqn.(\ref{gil-org}), cannot be ignored over long times. The first term corrects the mean--field acceleration of a star by a small amount, from $\bfa$ to $(1-1/N)\bfa$, taking note of the fact that the gravitational force on a star is due only to the other $(N-1)$ stars. It is the second term that is of importance for collisional evolution, where $f^{(2)}_{\rm irr}$ acts as the driving term. 

The equation for $f^{(2)}_{\rm irr}(\Gamma, \Gamma', t)$ depends on both $f(\Gamma, t)$ and $f(\Gamma', t)$, and is difficult to deal with in the general case. However, we are interested in describing the quasi--steady collisional relaxation of $f$. The stellar system can be thought of as evolving through a series of collisionless equilibria, for which $\left\{\bfv\cendot\p/\p\bfx + \left((1-1/N)\bfa + \bfa^{\rm e}\right)\cendot\p/\p\bfv\right\}\!f\simeq 0$, being of order $1/N$ or smaller. This implies that $\p f\!/\p t$ is first order in $1/N$, due solely to the collision term. It turns out that, in this case, $f^{(2)}_{\rm irr}$ can be expressed in terms of a conditional probability function $\scrw(\Gamma\,\vert\,\Gamma', t)$, which will henceforth be referred to as the Wake of $\Gamma'$ at $\Gamma$. The wake can be defined through the following \emph{gedanken} experiment: from the $N$ stars that are distributed in 6--dim phase space according to $f$, select one star and place it at the phase space location $\Gamma'$; this will induce a small perturbation $(1/N)\scrw$ at every location 
$\Gamma$ in the entire phase space. At this point it is useful to define two kinds of acceleration at any point $\bfx$. The wake acceleration, 
\beq
\bfa^{\rm w}(\bfx,\Gamma', t)
\;=\;\int \bfb(\bfx,\bfx'')
\scrw(\Gamma''\,\vert\,\Gamma', t)\,\rmd\Gamma''\,,
\label{a-wake}
\eeq
is the acceleration at $\bfx$ due to the wake of $\Gamma'$. 
We also define the perturbing acceleration,
\beq
\bfa^{\rm p}(\bfx, \bfx', t) \;=\; \bfb(\bfx, \bfx') \;-\; \bfa(\bfx, t)\,,
\label{pert-acc}
\eeq  
where the mean--field acceleration is subtracted from the bare inter--particle acceleration. The wake $\scrw(\Gamma\,\vert\,\Gamma', t)$ must be calculated by integrating an equation for the related wake function $\scrw(\Gamma\,\vert\,\Gamma'(t'), t')$, where $\Gamma'(t')=\left(\bfx'(t'), \bfv'(t')\right)$ is the location at time $t' \leq t$, of the star that arrives at the desired location $\Gamma'$ at time $t$. 
The equation for $\scrw(\Gamma\,\vert\,\Gamma'(t'), t')$ is:
\begin{eqnarray}
\frac{\p \scrw}{\p t'} \;+\; \bfv\cendot \frac{\p \scrw}{\p \bfx}
\;+\; \left\{\bfa(\bfx, t') \,+\, \bfa^{\rm e}(\bfx, t')\right\}\cendot\frac{\p\scrw}{\p \bfv}
\;+\; \bfa^{\rm w}(\bfx,\Gamma'(t'), t')\cendot\frac{\p f}{\p \bfv}&&\nonumber\\[1em] 
\qquad\qquad\qquad\;=\; -\bfa^{\rm p}(\bfx, \bfx'(t'), t')\cendot\frac{\p f}{\p \bfv}\,,\qquad\quad\mbox{for $t'\;\leq\; t\,$}.&&
\label{eqn-gil-wake}
\end{eqnarray}
Eqn.(\ref{eqn-gil-wake}) needs to be solved with the adiabatic turn--on initial condition $\scrw(\Gamma\,\vert\,\Gamma'(t'), t')\to 0$ as $t'\to -\infty$, corresponding to a vanishingly small wake in the distant past. 
The right hand side identifies the perturbing acceleration  $\bfa^{\rm p}(\bfx, \bfx'(t'), t')$ as the driver of the wake. The wake $\scrw(\Gamma\,\vert\,\Gamma', t)$ can be obtained by (formally) integrating eqn.(\ref{eqn-gil-wake}) from $t'=-\infty$ to $t'=t$, and is therefore a functional of the quasi--steady $f$. It may be verified that eqn.(\ref{eqn-gil-wake}) preserves $\int \scrw(\Gamma\,\vert\,\Gamma', t)\,\rmd\Gamma = 0$, because the wake contains no net mass.

Extending the work of \citet{ros64} in plasma physics to the case of an inhomogeneous stellar system, \citet{gil68} used non--trivial operator identities to derive the fundamental relation between $f^{(2)}_{\rm irr}$ and $\scrw$:
\begin{eqnarray}
f^{(2)}_{\rm irr}(\Gamma, \Gamma', t) &\;=\;& 
\scrw(\Gamma\,\vert\,\Gamma', t)\,f(\Gamma', t) \;+\;
\scrw(\Gamma'\,\vert\,\Gamma, t)\,f(\Gamma, t) 
\nonumber\\[1ex]
&&\qquad\qquad\;+\; \int \scrw(\Gamma\,\vert\,\Gamma'', t)\,
\scrw(\Gamma'\,\vert\,\Gamma'', t)\,
f(\Gamma'', t)\,\rmd\Gamma''\,.
\label{rostoker}
\end{eqnarray}
This decomposition of $f^{(2)}_{\rm irr}$ means that the irreducible 2--particle correlation at $(\Gamma, \Gamma')$ gets three contributions from
the wake function: (a) The wake of $\Gamma'$ at $\Gamma$; (b) The wake of 
$\Gamma$ at $\Gamma'$; (c) The product of the wake values at the points 
$\Gamma$ and $\Gamma'$ of a third Ring at $\Gamma''$, summed over all locations $\scrr''$. All three contributions come with suitable $f$--weighting. The third term accounts for the contribution of collective effects (``gravitational polarization'') to the microscopic processes driving RR. This is the nature of the full theory at $O(1/N)$. Since 
$\int f^{(2)}_{\rm irr}(\Gamma, \Gamma', t)\rmd\Gamma' = 0$, eqn.(\ref{rostoker}) implies that $\int \scrw(\Gamma\,\vert\,\Gamma', t)\,f(\Gamma', t)\rmd\Gamma' = 0$. 

Together, eqns.(\ref{gil-org}) and (\ref{rostoker}), give a closed set of kinetic equations governing the collisional relaxation of $f(\Gamma, t)$. This can also be written more explicitly as:
\begin{subequations}
\begin{eqnarray}
&&\frac{\p f}{\p t} \;+\; \bfv \cendot \frac{\p f}{\p \bfx}
\;+\; \left\{\left(1-\frac{1}{N}\right)\bfa + \bfa^{\rm ext}\right\}\cendot\frac{\p f}{\p \bfv} 
\;=\; \scrc^{\rm dis}\!\left[f\right] 
\;+\; \scrc^{\rm fluc}\!\left[f\right]\,,
\label{eqn-gke}\\[1em]
&&\scrc^{\rm dis}\!\left[f\right] \;=\; -\frac{1}{N}\int \bfb(\bfx,\bfx')\cendot
\frac{\p}{\p \bfv}\left\{f(\Gamma, t)\scrw(\Gamma'\,\vert\,\Gamma, t)
\right\}\,\rmd\Gamma'\,, 
\label{coll-dis}\\[1em]
&&\scrc^{\rm fluc}\!\left[f\right] \;=\; -\frac{1}{N}\int f(\Gamma', t)\left\{\bfb(\bfx,\bfx') + \bfa^{\rm w}(\bfx,\Gamma', t)\right\}\cendot\frac{\p \scrw(\Gamma\,\vert\,\Gamma', t)}{\p \bfv}\,\rmd\Gamma'\,. 
\label{coll-fluc}
\end{eqnarray}
\end{subequations}
Here $\scrc^{\rm dis}\!\left[f\right]$ and $\scrc^{\rm fluc}\!\left[f\right]$ are the dissipative and fluctuating parts of the collision term. As the notation indicates both are functionals of $f$, because they depend on the wake $\scrw(\Gamma\,\vert\,\Gamma', t)$ which, as discussed earlier, can be thought of as a functional of $f$.

\subsection{Gilbert's equations in Poisson Bracket form}

Gilbert's kinetic theory is expressed in the physically direct language 
of the various forces acting on a star. However, it is more convenient 
to exchange accelerations for corresponding potentials, and rewrite
the kinetic equations in Poisson Bracket (PB) form. We have, 
 \beq
\{\bfb, \,\bfa, \,\bfa^{\rm e}, \,\bfa^{\rm w}, \,\bfa^{\rm p}\} \;=\;
-\frac{\p}{\p\bfx}\{P, \,\Pi, \,\Pi^{\rm e}, \,\Pi^{\rm w}, \,\Pi^{\rm p}\}\,,
\label{acc-pot}
\eeq
where the potential functions on the right hand side are defined by:
\begin{subequations} 
\begin{eqnarray}
P(\bfx, \bfx') &\;=\;& -\frac{GM}{|\bfx -\bfx'|}\,,\quad\qquad\mbox{ 
(scaled) Inter--particle Poisson kernel;}
\label{int-part}\\[1ex]
\Pi(\bfx, t) &\;=\;& \int f(\Gamma', t)P(\bfx, \bfx')\,\rmd\Gamma'
\,,\quad\qquad\mbox{Mean field potential of the cluster;}
\label{mean-pot}\\[1ex]
\Pi^{\rm e}(\bfx, t) &\;=\;&
\mbox{External potential;}
\label{ext-pot}\\[1ex]
\Pi^{\rm w}(\bfx, \Gamma', t) &\;=\;& \int P(\bfx, \bfx'')\scrw(\Gamma''\,\vert\,\Gamma', t)\,\rmd\Gamma''
\,,\quad\qquad\mbox{Wake potential at $\bfx$ due to $\Gamma'$;}
\label{wake-pot}\\[1ex]
\Pi^{\rm p}(\bfx, \bfx', t) &\;=\;& P(\bfx, \bfx') \;-\; \Pi(\bfx, t)
\,,\quad\qquad\mbox{Perturbing potential.}
\label{pert-pot}
\end{eqnarray}
\end{subequations}
The mean--field Hamiltonian is:
\beq
H^{(1)}(\Gamma, t) \;=\; H^{(1)}(\bfx, \bfv, t) \;=\; 
\frac{v^2}{2} \;+\; \Pi(\bfx, t) \;+\;\Pi^{\rm e}(\bfx, t)\,,
\label{mf-ham}
\eeq
with the corresponding mean--field equations of motion,
\beq
\frac{\rmd\bfx}{\rmd t} \;=\; \frac{\p H^{(1)}}{\p \bfv} \;=\; \bfv\,,
\qquad \frac{\rmd\bfv}{\rmd t} \;=\; -\frac{\p H^{(1)}}{\p \bfx} \;=\;
\bfa \;+\ \bfa^{\rm e}\,.
\eeq

The PB between two phase space functions 
$\chi_1(\Gamma, t)$ and $\chi_2(\Gamma, t)$ is: 
\beq
\left[\,\chi_1\,,\,\chi_2\,\right]_{(\Gamma)} \;\stackrel{\rm def}{=}\;
\frac{\p\chi_1}{\p\bfx}\cendot\frac{\p\chi_2}{\p\bfv} \;-\;
\frac{\p\chi_1}{\p\bfv}\cendot\frac{\p\chi_2}{\p\bfx}\,.
\label{pb-xv}
\eeq
Using this we can rewrite eqn.(\ref{eqn-gil-wake}) for the wake function
$\scrw(\Gamma\,\vert\,\Gamma'(t'), t')$ as:
\begin{eqnarray}
&&\frac{\p \scrw}{\p t'} \;+\; \left[\,\scrw\,,\,H^{(1)}(\Gamma, t')\,\right]_{(\Gamma)}
\;+\; \left[\,f(\Gamma, t')\,,\,\Pi^{\rm w}(\bfx, \Gamma'(t'), t')\,\right]_{(\Gamma)}\nonumber\\[1ex] 
&&\qquad\qquad\;=\;  \left[\,\Pi^{\rm p}(\bfx, \bfx'(t'), t')\,,\,f(\Gamma, t')\,\right]_{(\Gamma)}\,,\qquad\quad\mbox{for $t'\;\leq\; t$,}
\nonumber\\[1ex]
&&\mbox{with adiabatic turn--on initial condition}\quad\lim_{t'\to -\infty}\scrw(\Gamma\,\vert\,\Gamma'(t'), t') \;=\; 0\,.
\label{eqn-gil-wake-pb}
\end{eqnarray}

There are two ways to write the Gilbert kinetic equations in PB form. 
The first is based on eqns.(\ref{gil-org}) and (\ref{rostoker}), which 
is closer to the original BBGKY equations: 
\begin{eqnarray}
&&\frac{\p f}{\p t} \;+\; \left[\,f(\Gamma, t)\,,\,H^{(1)}(\Gamma, t) - \frac{\Pi(\bfx, t)}{N}\,\right]_{(\Gamma)} 
\;=\; \frac{1}{N}\int \left[\,P(\bfx, \bfx')\,,\, f^{(2)}_{\rm irr}(\Gamma, \Gamma', t)\,\right]_{(\Gamma)}\,\rmd\Gamma'\,,\nonumber\\[1em]
&&f^{(2)}_{\rm irr}(\Gamma, \Gamma', t) \;=\; 
\scrw(\Gamma\,\vert\,\Gamma', t)\,f(\Gamma', t) \;+\;
\scrw(\Gamma'\,\vert\,\Gamma, t)\,f(\Gamma, t) 
\nonumber\\[1ex]
&&\qquad\qquad\qquad\qquad\qquad\;+\; \int \scrw(\Gamma\,\vert\,\Gamma'', t)\,\scrw(\Gamma'\,\vert\,\Gamma'', t)\,
f(\Gamma'', t)\,\rmd\Gamma''\,.
\label{bbgky-gil}
\end{eqnarray}
The second is the explicit form given by Gilbert, displaying the 
dissipative and fluctuating contributions to the collision integral.
From eqns.(\ref{eqn-gke})--(\ref{coll-fluc}) we have:
\begin{subequations}
\begin{eqnarray}
&&\frac{\p f}{\p t} \;+\; \left[\,f(\Gamma, t)\,,\,H^{(1)}(\Gamma, t) - \frac{\Pi(\bfx, t)}{N}\,\right]_{(\Gamma)} 
\;=\; \scrc^{\rm dis}\!\left[f\right] 
\;+\; \scrc^{\rm fluc}\!\left[f\right]\,,
\label{eqn-gke-pb}\\[1em]
&&\scrc^{\rm dis}\!\left[f\right] \;=\; \frac{1}{N}\int 
\left[\,P(\bfx, \bfx')\,,\,f(\Gamma, t)\scrw(\Gamma'\,\vert\,\Gamma, t)\,\right]_{(\Gamma)}\,\rmd\Gamma'\,,
\label{coll-dis-pb}\\[1em]
&&\scrc^{\rm fluc}\!\left[f\right] \;=\; \frac{1}{N}\int f(\Gamma', t)\left[\,P(\bfx, \bfx') + \Pi^{\rm w}(\bfx, \Gamma', t)\,,\,\scrw(\Gamma\,\vert\,\Gamma', t)\,\right]_{(\Gamma)}
\,\rmd\Gamma'\,. 
\label{coll-fluc-pb}
\end{eqnarray}
\end{subequations}
Either form of the kinetic equations applies to the collisional evolution of 
$f$ whenever it satisfies
\beq
\left[\,f(\Gamma, t)\,,\,H^{(1)}(\Gamma, t) - \frac{\Pi(\bfx, t)}{N}\,\right]_{(\Gamma)} \;\simeq\; 0\,,\quad\mbox{being of order $1/N$ or smaller.}
\label{qstead-gil}
\eeq
Therefore $f(\Gamma, t)$ may be thought of as passing through a sequence of collisionless equilibria in a quasi--static manner.

\section{Kinetic equation for a Stellar System orbiting a Massive
Black Hole}

We now specialize the kinetic theory to a stellar system of 
total mass $M=Nm_\star$, orbiting a MBH of mass $M_\bullet$. The mass ratio
$\varepsilon = M/M_\bullet$ is a natural ordering parameter; the smaller it is the more ``Keplerian'' is the stellar system. We begin with the 
kinetic equation in the BBGKY form, eqn.(\ref{bbgky-gil}), and make two canonical transformations. The first transformation is from the inertial frame to a (generally non--inertial) frame centered on the MBH. Then we transform to the Delaunay variables, which are natural action--angle variables for the Kepler problem. The formulae we derive are valid for arbitrary $\varepsilon$; however the second transformation proves really useful only when $\varepsilon$ is small.

\subsection{Transformation to coordinates centered on the MBH}

The Keplerian force of the MBH on a star is described by taking the 
external potential of eqn.(\ref{ext-pot}) as:
\beq
\Pi^{\rm e}(\bfx, t) \;=\;  -\frac{GM_\bullet}
{\left\vert\bfx \,-\, \bfx_\bullet(t)\right\vert}\,,
\label{bh-pot}
\eeq
where $\bfx_\bullet(t)$ is the position of the MBH with respect to the 
inertial frame. It is natural to transform to a non--inertial frame centered on the MBH. Let $\,\bfr = \bfx - \bfx_\bullet(t)\,$ and $\,\bfu = \bfv - \dot\bfx_\bullet(t)\,$ be the position and velocity, respectively, of a test star relative to the MBH. It is evident that $(\bfr, \bfu)$ are a canonically conjugate pair of 6--dim phase space coordinates. Using notation $\Gamma_\star = (\bfr, \bfu)$ we have $\rmd\Gamma \,=\, \rmd^3\bfx\rmd^3\bfv \,=\, \rmd^3\bfr\rmd^3\bfu \,=\, \rmd\Gamma_\star$. We write the (1--particle)
DF as $f(\Gamma_\star, t)$ and the wake as $\scrw(\Gamma_\star\,\vert\,\Gamma_\star', t)$. We take as an $O(1)$ reference the Kepler potential of the MBH potential $\Pi^{\rm ext} = -GM_\bullet/r\,$ given in eqn.(\ref{bh-pot}). Then the other potentials in eqns.(\ref{int-part}), (\ref{mean-pot}), (\ref{wake-pot}) and (\ref{pert-pot}) are all quantities of $O(\varepsilon)$. Hence we rescale these and write $\{P, \Pi, \Pi^{\rm w}, \Pi^{\rm p}\} \,=\, \varepsilon\{p, \varphi, \varphi^{\rm w}, \varphi^{\rm p}\}$, where 
\begin{subequations} 
\begin{eqnarray}
p(\bfr, \bfr') &\;=\;& -\frac{GM_\bullet}{|\bfr -\bfr'|}\,,\quad\qquad\mbox{(rescaled) Poisson kernel;}
\label{int-part-scal}\\[1ex]
\varphi(\bfr, t) &\;=\;& \int f(\Gamma_\star', t)p(\bfr, \bfr')\,\rmd\Gamma_\star'\,,\quad\qquad\mbox{(rescaled) Mean field potential;}
\label{mean-pot-scal}\\[1ex]
\varphi^{\rm w}(\bfr, \Gamma_\star', t) &\;=\;& \int p(\bfr, \bfr'')\scrw(\Gamma_\star''\,\vert\,\Gamma_\star', t)\rmd\Gamma_\star''
\,,\quad\qquad\mbox{(rescaled) Wake potential;}
\label{wake-pot-scal}\\[1ex]
\varphi^{\rm p}(\bfr, \bfr', t) &\;=\;& p(\bfr, \bfr') \;-\; \varphi(\bfr, t)
\,,\quad\qquad\mbox{(rescaled) Perturbing potential.}
\label{pert-pot-scal}
\end{eqnarray}
\end{subequations}

All the Poisson Brackets in eqn.(\ref{eqn-gil-wake-pb}) and eqns.(\ref{eqn-gke-pb})--(\ref{coll-fluc-pb}) remain invariant, because $(\bfx, \bfv) \to (\bfr, \bfu)$ is a canonical transformation. i.e. for any two phase space functions $\chi_1$ and $\chi_2$, the PB defined earlier in eqn.(\ref{pb-xv}) is: 
\beq
\left[\,\chi_1\,,\,\chi_2\,\right]_{(\Gamma)} \;=\;
\left[\,\chi_1\,,\,\chi_2\,\right]_{(\Gamma_\star)} \;\;\stackrel{\rm def}{=}\;\;
\frac{\p\chi_1}{\p\bfr}\cendot\frac{\p\chi_2}{\p\bfu} \;-\;
\frac{\p\chi_1}{\p\bfu}\cendot\frac{\p\chi_2}{\p\bfr}\,.
\label{pb-ru}
\eeq
The first term on the left hand side of the wake eqn.(\ref{eqn-gil-wake-pb})
is $(\p\scrw/\p t')_{(\Gamma)}$ where we have now indicated explicitly that $\Gamma=(\bfx, \bfv)$ is to be held constant. We need to express this in terms of functions of $\Gamma_\star = (\bfr, \bfu)$ and $t$. The chain rule gives,
\begin{eqnarray}
\left(\frac{\p \scrw}{\p t'}\right)_{(\Gamma)} &\;=\;&
\left(\frac{\p \scrw}{\p t'}\right)_{(\Gamma_\star)} \;-\;
\dot\bfx_\bullet\cendot\frac{\p \scrw}{\p \bfr} \;-\;
\ddot\bfx_\bullet\cendot\frac{\p \scrw}{\p \bfu}\nonumber\\[1em]
&\;=\;& \left(\frac{\p \scrw}{\p t'}\right)_{(\Gamma_\star)} \;+\; 
\left[\,\scrw\,,\;\bfr\cendot\ddot\bfx_\bullet(t) - \bfu\cendot\dot\bfx_\bullet(t)
-\frac{1}{2}\vert\dot\bfx_\bullet(t)\vert^2\,\right]_{(\Gamma_\star)}
\label{time-der}
\end{eqnarray}
Then eqn.(\ref{eqn-gil-wake-pb}) for the wake function $\scrw(\Gamma_\star\,\vert\,\Gamma_\star'(t'), t')$ is:
\begin{eqnarray}
&&\frac{\p \scrw}{\p t'} \;+\; \left[\,\scrw\,,\,H_{\rm org}(\Gamma_\star, t')\,\right]_{(\Gamma_\star)}
\;+\; \varepsilon\left[\,f(\Gamma_\star, t')\,,\,\varphi^{\rm w}(\bfr, \Gamma_\star'(t'), t')\,\right]_{(\Gamma_\star)}\nonumber\\[1em] 
&&\qquad\qquad\;=\;  \varepsilon\left[\,\varphi^{\rm p}(\bfr, \bfr'(t'), t')\,,\,f(\Gamma_\star, t')\,\right]_{(\Gamma_\star)}\,,\qquad\quad\mbox{for $t'\;\leq\; t$,}
\nonumber\\[1em]
&&\mbox{with adiabatic turn--on initial condition}\quad\lim_{t'\to -\infty}\scrw(\Gamma_\star\,\vert\,\Gamma_\star'(t'), t') \;=\; 0\,.
\label{eqn-gil-wake-ru}
\end{eqnarray}
We have written $H_{\rm org}(\bfr, \bfu, t) \,=\, H^{(1)} + \{\bfr\cendot\ddot\bfx_\bullet(t) - \bfu\cendot\dot\bfx_\bullet(t) -\vert\dot\bfx_\bullet(t)\vert^2/2\}\,$, and dropped the subscript $\,\Gamma_\star$ in the time derivative. The acceleration of the MBH, $\,\ddot\bfx_\bullet(t)\,$, is due to the gravitational forces of all the stars, and is a $O(\varepsilon)$ quantity. Rescaling, we write $\ddot\bfx_\bullet(t) = \varepsilon\bfA_\bullet(t)$, where
\beq
\bfA_\bullet(t) \;=\; GM_\bullet\int 
f(\Gamma_\star, t)\,\frac{\hat{\bfr}\;}{r^2}
\,\rmd\Gamma_\star\,.
\label{aself}
\eeq
Using eqn.(\ref{mf-ham}) for $H^{(1)}$ we can write 
\beq
H_{\rm org}(\Gamma_\star, t) \;=\; H_{\rm org}(\bfr, \bfu, t) \;=\; \frac{u^2}{2} \,-\, \frac{GM_\bullet}{r} \;+\;
\varepsilon\varphi(\bfr, t) \;+\; 
\varepsilon\bfr\cendot\bfA_\bullet(t)
\label{ham0}
\eeq
explicitly as a function of $(\bfr, \bfu, t)$. This is the Hamiltonian governing the mean--field dynamics of $\bfr$ and $\bfu$ through the 
equations of motion:  
\beq
\frac{\rmd\bfr}{\rmd t} \;=\; \frac{\p H_{\rm org}}{\p \bfu} \;=\; \bfu\,,
\qquad \frac{\rmd\bfu}{\rmd t} \;=\; -\frac{\p H_{\rm org}}{\p \bfr} \;=\;
-\,\frac{GM_\bullet}{r^2}\hat{\bfr} \;-\; \varepsilon\frac{\p \varphi}{\p \bfr} \;-\; \varepsilon\bfA_\bullet\,,
\eeq
as can be verified from the definitions of $\bfr$ and $\bfu$. Note that 
both $H_{\rm org}$ and the equations of motion are identical to those 
given in \S~2 of Paper~I. Similarly, eqn.(\ref{bbgky-gil}) can be rewritten in the MBH variables. Then $f(\Gamma_\star, t)$ satisfies:
\begin{eqnarray}
&&\frac{\p f}{\p t} \;+\; \left[\,f(\Gamma_\star, t)\,,\,H_{\rm org}(\Gamma_\star, t) - \frac{\varepsilon}{N}\varphi(\bfr, t)\,\right]_{(\Gamma_\star)} 
\;=\; 
\frac{\varepsilon}{N}\int \left[\,p(\bfr, \bfr')\,,\, 
f^{(2)}_{\rm irr}(\Gamma_\star, \Gamma_\star', t)\,\right]_{(\Gamma_\star)}\,\rmd\Gamma_\star'\,,\nonumber\\[1em]
&&f^{(2)}_{\rm irr}(\Gamma_\star, \Gamma_\star', t) \;=\; 
\scrw(\Gamma_\star\,\vert\,\Gamma_\star', t)\,f(\Gamma_\star', t) \;+\;
\scrw(\Gamma_\star'\,\vert\,\Gamma_\star, t)\,f(\Gamma_\star, t) 
\nonumber\\[1ex]
&&\qquad\qquad\qquad\qquad\qquad\;+\; \int \scrw(\Gamma_\star\,\vert\,\Gamma_\star'', t)\,\scrw(\Gamma_\star'\,\vert\,\Gamma_\star'', t)\,
f(\Gamma_\star'', t)\,\rmd\Gamma_\star''\,.
\label{bbgky-gil-ru}
\end{eqnarray}

\subsection{Transformation to the Delaunay variables}

We now transform from the MBH--centric canonical coordinates, $\Gamma_\star = (\bfr, \bfu)$, to the Delaunay variables $\scrd\equiv\{I, L, L_z; w, g, h\}$, which are action--angle variables for the Kepler problem \citep{plu60,md99,bt08}. The three actions are: $I\,=\,\sqrt{GM_\bullet a\,}\,$; $L\,=\, I\sqrt{1-e^2\,}$ the magnitude of the angular momentum; and $L_z\,=\, L\cos{i}\,$ the $z$--component of the angular momentum. The angles conjugate to them are, respectively: $w$ the orbital phase; $g$ the argument of the periapse; and $h$ the longitude of the ascending node. 
The transformation is exact and no restriction is placed on the magnitude of 
$\varepsilon$ (although this is preparatory to the orbit--averaging that follows in the next section for $\varepsilon \ll 1$). In the Hamiltonian $H_{\rm org}$ of eqn.(\ref{ham0}), the Keplerian orbital energy $E_{\rm k} = \left(u^2/2 - GM_\bullet/r\right)\,$ will be taken as $O(1)$. The other two terms, $\varepsilon\varphi$ and $\varepsilon\bfr\cendot\bfA_\bullet$, 
are evidently of $O(\varepsilon)$.

The Kepler orbital energy $E_{\rm k}(I) = -1/2(GM_\bullet/I)^2$ depends only on the action $I$. Therefore, for the pure Kepler problem (i.e. when $\varepsilon = 0$) all the Delaunay variables, excepting $w$, are constant in time; $w$ itself advances at the rate 
\beq
\Omega_{\rm k}(I) \;=\; \frac{{\rm d} E_{\rm k}}{{\rm d} I} 
\;=\; \frac{(GM_\bullet)^2}{I^3}\,.
\label{kepfreq}
\eeq
In order to rewrite the various physical quantities in terms of the 
$\scrd$ variables we need to express the position vector $\bfr(\scrd) = (x, y, z)$ in terms of the $\scrd$ variables \citep{plu60,md99,ss00}:
\beq
{\left( \begin{array}{c} x \\ \\ y \\ \\ z \end{array} \right) } 
 = 
{\left( \begin{array}{ccc} 
   C_{g} C_{h} - C_{i} S_{h} S_{g} & \quad -S_{g} C_{h} - C_{i} S_{h} C_{g} & 
   \quad S_{i} S_{h}  \\ \\
   C_{g} S_{h} + C_{i} C_{h} S_{g} & \quad -S_{g} S_{h} + C_{i} C_{h} C_{g} & 
   \quad -S_{i} C_{h} \\ \\
   S_{i} S_{g}               & \quad S_{i} C_{g}            & \quad C_{i}
	\end{array} \right)} {\left( \begin{array}{c} 
   a(C_{\eta} - e) \\  \\ a\sqrt{1 - e^2\,}\, S_{\eta} \\  \\ 0 \end{array} 
   \right)}
\label{xyzdel}
\eeq
\noindent 
where $S$ and $C$ are shorthand for sine and cosine of the angles given as 
subscript. Here $a$ is the semi--major axis; $\eta$ is the eccentric anomaly, related to the orbital phase through $w= (\eta - e\sin\eta)\,$; 
$\,e=\sqrt{1-L^2/I^2\,}$ is the eccentricity; and $i$ is the inclination angle determined by $\cos i = (L_z/L)\,$. It is also useful to note that
$r = \sqrt{x^2 + y^2 + z^2} = a(1 - e\cos\eta)\,$. 

The DF is written as $f(\scrd, t)$ and the wake as $\scrw(\scrd\,\vert\,\scrd', t)$. The transformation $\Gamma_\star \to \scrd$ being canonical,
$\rmd\Gamma_\star = \rmd\scrd$. Then the potentials of eqn.(\ref{int-part-scal})--(\ref{pert-pot-scal}) and $\bfA_\bullet$ can 
be written as:
\begin{subequations} 
\begin{eqnarray}
p(\scrd, \scrd') &\;=\;& -\frac{GM_\bullet}{|\bfr -\bfr'|}\,,\quad\qquad\mbox{(rescaled) Poisson kernel;}
\label{int-part-del}\\[1ex]
\varphi(\scrd, t) &\;=\;& \int f(\scrd', t)p(\scrd, \scrd')\,\rmd\scrd'\,,\quad\qquad\mbox{(rescaled) Mean field potential;}
\label{mean-pot-del}\\[1ex]
\bfA_\bullet(t) &\;=\;& GM_\bullet\int f(\scrd, t)\,\frac{\hat{\bfr}\;}{r^2}
\,{\rm d}\scrd\,,\quad\qquad\mbox{(rescaled) MBH acceleration;}
\label{accn-del}\\[1ex]
\varphi^{\rm w}(\scrd, \scrd', t) &\;=\;& \int p(\scrd, \scrd'')\scrw(\scrd''\,\vert\,\scrd', t)\,\rmd\scrd''
\,,\quad\qquad\mbox{(rescaled) Wake potential;}
\label{wake-pot-del}\\[1ex]
\varphi^{\rm p}(\scrd, \scrd', t) &\;=\;& p(\scrd, \scrd') \;-\; \varphi(\scrd, t)
\,,\quad\qquad\mbox{(rescaled) Perturbing potential.}
\label{pert-pot-del}
\end{eqnarray}
\end{subequations}
The mean--field Hamiltonian is: 
\beq
H_{\rm org}(\scrd, t) \;=\; E_{\rm k}(I) \;+\; \varepsilon\varphi(\scrd, t)
\;+\; \varepsilon\bfr\cendot\bfA_\bullet(t)\,.
\label{ham0-del}
\eeq
Since the transformation $\Gamma_\star \to \scrd$ is canonical, 
all the PBs remain invariant. The PB between any two phase space 
functions $\chi_1(\scrd)$ and $\chi_2(\scrd)\,$ is:
\beq
\left[\,\chi_1\,,\,\chi_2\,\right]_{(6)} \stackrel{\rm def}{=}
\left(\frac{\p \chi_1}{\p w}\frac{\p \chi_2}{\p I} -
\frac{\p \chi_1}{\p I}\frac{\p \chi_2}{\p w}\right) \,+\, 
\left(\frac{\p \chi_1}{\p g}\frac{\p \chi_2}{\p L} -
\frac{\p \chi_1}{\p L}\frac{\p \chi_2}{\p g}\right) \,+\, 
\left(\frac{\p \chi_1}{\p h}\frac{\p \chi_2}{\p L_z} -
\frac{\p \chi_1}{\p L_z}\frac{\p \chi_2}{\p h}\right)\,.
\label{pbdel}
\eeq
From eqn.(\ref{eqn-gil-wake-ru}), we see that the wake function $\scrw(\scrd\,\vert\,\scrd'(t'), t')$ satisfies
\begin{eqnarray}
&&\frac{\p \scrw}{\p t'} \;+\; \Omega_{\rm k}\frac{\p \scrw}{\p w} \;+\;
\varepsilon\left[\,\scrw\,,\,\varphi \,+\, \bfr\cendot\bfA_\bullet\,\right]_{(6)} \;+\; \varepsilon\left[\,f(\scrd, t')\,,\,\varphi^{\rm w}(\scrd, \scrd'(t'), t')\,\right]_{(6)}\nonumber\\[1em]
&&\qquad\qquad\;=\;  \varepsilon\left[\,\varphi^{\rm p}(\scrd, \scrd'(t'), t')\,,\,f(\scrd, t')\,\right]_{(6)}\,,\qquad\quad\mbox{for $t'\;\leq\; t$,}
\nonumber\\[1em]
&&\mbox{with adiabatic turn--on initial condition}\quad\lim_{t'\to -\infty}\scrw(\scrd\,\vert\,\scrd'(t'), t') \;=\; 0\,.
\label{eqn-gil-wake-del}
\end{eqnarray}
Similarly, from eqn.(\ref{bbgky-gil-ru}), the
kinetic equation for $f(\scrd, t)$ is:
\begin{eqnarray}
&&\frac{\p f}{\p t} \;+\; \Omega_{\rm k}\frac{\p f}{\p w} \;+\;
\varepsilon\left[\,f\,,\,\left(1 - \frac{1}{N}\right)\varphi \,+\, \bfr\cendot\bfA_\bullet\,\right]_{(6)} \;=\; \nonumber\\[1ex]
&&\qquad\qquad\qquad\qquad\qquad\frac{\varepsilon}{N}\int \left[\,p(\scrd, \scrd')\,,\, f^{(2)}_{\rm irr}(\scrd, \scrd', t)\,\right]_{(6)}\,\rmd\scrd'\,,\nonumber\\[1em]
&&f^{(2)}_{\rm irr}(\scrd, \scrd', t) \;=\; 
\scrw(\scrd\,\vert\,\scrd', t)\,f(\scrd', t) \;+\;
\scrw(\scrd'\,\vert\,\scrd, t)\,f(\scrd, t) 
\nonumber\\[1ex]
&&\qquad\qquad\qquad\qquad\qquad\;+\; \int \scrw(\scrd\,\vert\,\scrd'', t)\,\scrw(\scrd'\,\vert\,\scrd'', t)\,
f(\scrd'', t)\,\rmd\scrd''\,.
\label{bbgky-gil-del}
\end{eqnarray}

\section{Orbit--averaging for a Keplerian Stellar System}

The stellar system is said to be Keplerian when the mass ratio 
$\varepsilon = M/M_\bullet \ll 1$. In this case the dominant force on 
a star is the inverse--squared Newtonian force of the MBH, and 
$\varepsilon$ is the natural small parameter for a perturbation theory
of Keplerian systems. We want to simplify the description of the previous section and obtain an $O(\varepsilon)$ description of the quasi--steady collisional evolution.
 
\subsection{Elements of Secular Collisionless dynamics}

The first step is to understand the general features of the collisionless limit, $N\to\infty\,$ and  $\,m_\star \to 0\,$ with 
$\,M=\mbox{constant}$, of the kinetic eqn.(\ref{bbgky-gil-del}) for 
$f(\scrd, t)$:
\beq
\frac{\p f}{\p t} \;+\; \Omega_{\rm k}\frac{\p f}{\p w} \;+\;
\varepsilon\left[\,f\,,\,\varphi\,\right]_{(6)} \;=\; 0\,.
\label{cbe-full}
\eeq
We begin by recalling results from the theory presented in Paper~I,
which is accurate to $O(\varepsilon)$. The natural time scales in the problem are (a) the Kepler orbital period $\tk = 2\pi/\Omega_{\rm k}\,$, on which the phase $w$ varies, and (b) the longer secular time scale 
$\ts = \varepsilon^{-1}\tk\,$ over which the other 5 Delaunay variables $\scrr\equiv\{I, L, L_z; g, h\}\,$ evolve in the mean.\footnote{The relevant time scales for the two best--known Keplerian systems, the Galactic Centre 
and the nucleus of M31, are discussed at the end of the introductory 
part of \S~2 in Paper~I.} The CBE describing secular evolution was derived by using the method of multiple scales, to orbit--average the full CBE over the fast orbital phase $w$. Two time variables $t$ and a slow time $\tau = \varepsilon t$ were introduced; these are natural measures of time over $\tk$ and $\ts$, respectively. We argued that the secular description cannot depend on the fast orbital phase $w$. The remaining 5 Delaunay variables $\scrr\equiv\{I, L, L_z; g, h\}\,$ are the coordinates of a Gaussian Ring (or just Ring) which is a Keplerian ellipse of given semi--major axis, eccentricity, inclination, periapse angle and nodal longitude. The secular evolution of $N\gg 1$ stars can be thought of as the evolution of $N\gg 1$ Gaussian Rings, so the $O(1)$ part of the DF must be a function of the slow variables $(\scrr, \tau)$, and independent of the fast variables $(w, t)$. It turns out that the appropriate expansion of the DF, to $O(\varepsilon)$ 
accuracy, is: 
\beq
f(\scrd, t, \tau) \;=\; \frac{1}{2\pi}\left\{F(\scrr, \tau) + \varepsilon
f_0(\scrr, \tau)\right\} \;+\; 
\frac{\varepsilon}{2\pi}\,\sum_{n\neq 0}f_n(\scrr, t, \tau)\exp{[{\rm i}n w]} \;+\; O(\varepsilon^2)\,, 
\label{df-pap1}
\eeq
where $F(\scrr, \tau)$ will be referred to as the Ring DF. 
The DF $f$ is a probability distribution function in $\scrd$--space satisfying $\int f\,\rmd\scrd = 1$ in the absence of loss of stars to the MBH --- see eqn.(1) of Paper~I. Using ${\rm d}\scrd = {\rm d}\scrr\,{\rm d}w\,$, we have
\beq
\int F(\scrr, \tau)\,\rmd\scrr \;+\; \varepsilon\int f_0(\scrr, \tau)\,\rmd\scrr  \;+\; O(\varepsilon^2) \;=\; 1\,.
\label{sec-norm}
\eeq
The method of multiple scales provided a solution to $O(\varepsilon)$.
We returned to the usual single--time description, and presented the following solution to the full CBE eqn.(\ref{cbe-full}):
\beq
f(\scrd, t) \;=\; \frac{1}{2\pi}\left\{F(\scrr, \varepsilon t) + \varepsilon f_0(\scrr, \varepsilon t)\right\} \;+\;
\frac{\varepsilon}{2\pi}\Delta F(\scrd, \varepsilon t) \;+\;
\frac{\varepsilon}{2\pi}\Lambda\!\left(\scrr, \,w- \Omega_{\rm k} t, \,\varepsilon t\right)
\;+\; O(\varepsilon^2)\,.  
\label{cbe-fullsoln}
\eeq
Here $F(\scrr, \tau)$ satisfies the Ring CBE: 
\beq
\frac{\p F}{\p \tau} \;+\; \left[\,F\,,\,\Phi\,\right] \;=\; 0\,,
\label{cbe-sec}
\eeq
where $\,\Phi(\scrr, \tau)$ is the Ring Hamiltonian (coming from stellar self--gravity), 
\beq
\Phi(\scrr, \tau) \;=\; \oint\varphi(\scrd, \tau)\,\frac{\rmd w}{2\pi}\,,
\label{ham-sg}
\eeq
which is equal to the orbit--averaged mean--field potential. The PB without subscript is the 4--dim PB, 
\beq
\left[\,\chi_1\,,\,\chi_2\,\right] \;=\; 
\left(\frac{\p \chi_1}{\p g}\frac{\p \chi_2}{\p L} -
\frac{\p \chi_1}{\p L}\frac{\p \chi_2}{\p g}\right) \,+\, 
\left(\frac{\p \chi_1}{\p h}\frac{\p \chi_2}{\p L_z} -
\frac{\p \chi_1}{\p L_z}\frac{\p \chi_2}{\p h}\right)\,,
\label{pbdel4}
\eeq
whose action is restricted to the 4--dim $I=\mbox{constant}\;$ surfaces in the 5--dim $\scrr$--space. Under the self--consistent Ring dynamics of eqn.(\ref{cbe-sec}), $F(\scrr, \tau)$ can have significant time variations over times $\ts$. 

There are three kinds of $O(\varepsilon)$ corrections to the Ring description. $\,f_0(\scrr, \tau)$ is a secular correction term to the Ring DF, which remains undetermined in our $O(\varepsilon)$ theory. The other 
two correction terms are purely fluctuating in $w$. Of these, $\,\Delta F(\scrd, \varepsilon t)$ is a known function of the Ring DF $F$ and $w$; 
every Ring DF $F(\scrr, \varepsilon t)$ in the 5--dim $\scrr$--space is accompanied by a small fluctuating (in $w$) distortion in the full 6--dim 
$\scrd$--space. $\,\Lambda$ is an arbitrary function of its arguments 
$\scrr$, $(w-\Omega_{\rm k} t)$ and $\varepsilon t$, with zero mean over 
$w$. The arbitrariness of form is due to the limits of our $O(\varepsilon)$
theory. Yet it has physical significance, in that it emphasizes the 
requirement that there are no instabilities growing on the fast orbital times $\tk$ --- as it must be, for $F(\scrr, \tau)$ to be a physical 
$O(1)$ description.  

The $O(1)$ self--consistent secular collisionless description is obtained by taking the limit $\varepsilon \to 0$, while the time scales of interest are large enough to make $\tau$ a sensible $O(1)$ measure of time. This limit is described completely by the Ring DF, $F(\scrr, \tau)$, which satisfies the Ring CBE (\ref{cbe-sec}). The normalization of eqn.(\ref{sec-norm}) reduces to 
\beq
\int F(\scrr, \tau)\,\rmd\scrr  \;=\; 1\,,
\label{Fnorm}
\eeq
so $F(\scrr, \tau)$ may be thought of as a probability distribution function in the 5--dim $\scrr$--space. The orbit of every Ring is restricted to its 4--dim $I=\mbox{constant}\;$ surface. The dynamics of the remaining 4 Ring variables $\{L, L_z; g, h\}$ is given by the Hamiltonian equations, 
\beq
\frac{\rmd L}{\rmd \tau} \;=\; - \frac{\p \Phi}{\p g}\,,\qquad
\frac{\rmd g}{\rmd \tau} \;=\;  \frac{\p \Phi}{\p L}\,;\qquad
\frac{\rmd L_z}{\rmd \tau} \;=\; - \frac{\p \Phi}{\p h}\,,\qquad
\frac{\rmd h}{\rmd \tau} \;=\;  \frac{\p \Phi}{\p L_z}\,.
\label{ring-eom-sg}
\eeq
which will be referred to as the Ring equations of motion. Using these, 
it is straightforward to write the the Ring CBE (\ref{cbe-sec})as ${\rm d} F/{\rm d} \tau \,=\, 0\,$, justifying the term ``collisionless''. The Ring description makes no reference to the orbital phase $w$, and is self-consistent and closed in 5--dim $\scrr$--space. However, we will later need to take account of the evolution of $w$ to $O(1)$. It turns out that this is slaved to the Ring degrees of freedom, and given by:
\beq
\frac{\rmd w}{\rmd \tau} \;=\; \frac{\Omega_{\rm k}(I)}{\varepsilon} \;+\; 
\frac{\p \Phi}{\p I}\,. 
\label{orb-rate-sg}
\eeq
The orbital phase increases steadily over times $\tk$, and also has modulations over times $\ts$.

\subsection{Orbit--averaged kinetic equation}

We now return to the problem of quasi--steady collisional evolution
of Keplerian stellar systems, and simplify the kinetic equation of the 
previous section through first order perturbation theory in $\varepsilon$.
In quasi--steady evolution it is clear that the DF $f$ cannot depend on $t$, and hence must be of the form $f(\scrd, \tau)$. Moreover, in the collisionless limit, the DF must go smoothly to the form given in eqn.(\ref{df-pap1}). Therefore we write:
\beq
f(\scrd, \tau) \;=\; \frac{1}{2\pi}F(\scrr, \tau) \;+\; 
\frac{\varepsilon}{2\pi}\,\sum_{n=-\infty}^{\infty}f_n(\scrr, \tau)\exp{[{\rm i}n w]} \;+\; O(\varepsilon^2)\,,
\label{dffou}
\eeq
where $F\geq 0$ and $f_{-n} = f_n^\star\,$. We recall from \S~2 that the wake is a ``response function'' characterizing the collisionless system which, to $O(1)$, is described completely by the secular DF $F(\scrr, \tau)$. Hence to $O(1)$ the wake must describe the perturbation in $\scrr$--space due to the selection of a Ring at a given location. Therefore the power--series in $\varepsilon$ for the wake can be written as:
\beq
\scrw(\scrd\,\vert\,\scrd', \tau) \;=\; \frac{1}{2\pi}W(\scrr\,\vert\,\scrr', \tau) \;+\; \frac{\varepsilon}{2\pi}\sum_{n=-\infty}^{\infty}\varpi_n(\scrr\,\vert\,\scrd', \tau)\exp{[{\rm i}n w]} \;+\; O(\varepsilon^2)\,,
\label{wake-gr}
\eeq
where $(1/N)W(\scrr\,\vert\,\scrr', \tau)$ can be thought of as the additional perturbation at $\scrr$, resulting from the selection of a 
Ring at $\scrr'$. We will refer to $W(\scrr\,\vert\,\scrr', \tau)$ as the 
Ring Wake function. 

The Gilbert equations, eqns.(\ref{eqn-gil-wake-del}) and (\ref{bbgky-gil-del}), are valid for any value of the mass ratio $\varepsilon$. Orbit--averaging is a systematic, perturbative method that can be applied to a Keplerian stellar system for which $\varepsilon \ll 1$. This proceeds by substituting the multiple--scale forms, eqn.(\ref{dffou}) for the DF $f$ and eqn.(\ref{wake-gr}) for the wake $\scrw$, in the Gilbert equations and developing them perturbatively in the small paramater $\varepsilon$. The
calculations are straightforward yet tedious, and we provide the details 
in the Appendix. Henceforth we discuss the $O(1)$ theory for the collisional evolution of $F(\scrr, \tau)$ and $W(\scrr\,\vert\,\scrr', \tau)$ in 5--dim 
$\scrr$--space --- the fluctuating quantities, $f_n$ and $\varpi_n$,  contribute only at higher order. In contrast to Gilbert's kinetic theory of point masses this is a kinetic theory of Gaussian Rings, and is the appropriate framework for describing the Resonant Relaxation of RT96. Analogous to the various potentials that arise in the PB form of Gilbert's equations, various Ring potentials make their appearance. These are all derived in the Appendix --- see eqns.(\ref{Psidef}), (\ref{phislow}), (\ref{phislow-w}) and (\ref{phislow-p}) --- and reproduced below: 
\beq
\Psi(\scrr, \scrr') \;=\; -GM_\bullet\oint\oint\frac{{\rm d}w}{2\pi}\,
\frac{{\rm d}w'}{2\pi}\,\frac{1}{\left|\bfr - \bfr'\right|}\,,
\nonumber
\eeq
is the ``bare'' inter--Ring potential in $\scrr$--space, and plays a role 
analogous to the Poisson kernel, $p(\scrd, \scrd')$, in Gilbert's theory. The other Ring potential are: 
\begin{eqnarray}
\Phi(\scrr, \tau) &\;=\;& \int F(\scrr', \tau)\,\Psi(\scrr, \scrr')
{\rm d}\scrr'\,,\qquad\quad\mbox{Ring mean--field potential;}
\nonumber\\[1ex]
\Phi^{\rm w}(\scrr, \scrr', \tau) &\;=\;& \int \,W(\scrr''\,\vert\,\scrr', \tau)\,\Psi(\scrr, \scrr'')\,{\rm d}\scrr''\,,\qquad\quad\mbox{Ring wake potential;}
\nonumber\\[1ex]
\Phi^{\rm p}(\scrr, \scrr', \tau) &\;=\;& \Psi(\scrr, \scrr') \;-\; \Phi(\scrr, \tau)\,,\qquad\quad\mbox{Ring perturbing potential.}
\nonumber 
\end{eqnarray}
All of these have the same significance as the corresponding potentials 
in Gilbert's theory, as is clear from their appearance in the governing 
equations for $W$ and $F$ given below. 
 
From eqn.(\ref{eqn-gil-wake-kepW3}) the wake function $W(\scrr\,\vert\,\scrr'(\tau'), \tau')$ is determined by:
\begin{eqnarray}
&&\frac{\p W}{\p \tau'} \;+\; \left[\,W(\scrr\,\vert\,\scrr'(\tau'), \tau')\,,\,\Phi(\scrr, \tau')\,\right]
\;+\; \left[\,F(\scrr, \tau')\,,\,\Phi^{\rm w}(\scrr, \scrr'(\tau'), \tau')\,\right]
\nonumber\\[1em]
&&\qquad\qquad \;=\; \left[\,\Phi^{\rm p}(\scrr, \scrr'(\tau'),\tau')\,,\,F(\scrr, \tau')\,\right]
\,,\qquad\quad\mbox{for $\tau'\;\leq\; \tau$,}
\nonumber\\[1em]
&&\mbox{with adiabatic turn--on initial condition}\quad\lim_{\tau'\to -\infty} W(\scrr\,\vert\,\scrr'(\tau'), \tau') \;=\; 0\,.
\label{eqn-gil-wake-kepW4}
\end{eqnarray}
From eqn.(\ref{bbgky-gil-kepF3}) and (\ref{irr-kep0}) the kinetic equation for the Ring DF $F(\scrr, \tau)$ is:
\begin{eqnarray}
&&\frac{\p F}{\p \tau} \;+\; 
\left(1 - \frac{1}{N}\right)\left[\,F\,,\,\Phi(\scrr, \tau)\,\right] \;=\; 
\frac{1}{N}\int\left[\,\Psi(\scrr, \scrr')\,,\, F^{(2)}_{\rm irr}(\scrr, \scrr', \tau)\,\right]\,\rmd\scrr'\,,
\nonumber\\[1em] 
&&F^{(2)}_{\rm irr}(\scrr, \scrr', \tau) \;=\; 
W(\scrr\,\vert\,\scrr', \tau)\,F(\scrr', \tau) \;+\;
W(\scrr'\,\vert\,\scrr, \tau)\,F(\scrr, \tau) 
\nonumber\\[1ex]
&&\qquad\qquad\qquad\qquad\qquad\;+\; \int W(\scrr\,\vert\,\scrr'', \tau)\,W(\scrr'\,\vert\,\scrr'', \tau)\,F(\scrr'', \tau)\,\rmd\scrr''\,.
\label{bbgky-gil-kepF4}
\end{eqnarray}
Eqns.(\ref{eqn-gil-wake-kepW4}) and (\ref{bbgky-gil-kepF4}) are the basic equations governing the RR of a non--relativistic and isolated Keplerian stellar system. In the Appendix we discuss the $O(\varepsilon)$ fluctuations and argue that they are bounded and hence eqns.(\ref{eqn-gil-wake-kepW4}) and (\ref{bbgky-gil-kepF4}) form a closed and consistent set of equations in 5--dim $\scrr$--space. An important feature of the evolution of $F(\scrr, \tau)$, as described by  eqns.(\ref{eqn-gil-wake-kepW4}) and (\ref{bbgky-gil-kepF4}), is the absence of the mass ratio $\varepsilon$ in all the physical quantities, except in the time variable $\tau = \varepsilon t$. As pointed out by RT96 this implies the following property of mass--invariance: if the mass of every star in a Keplerian cluster is changed from $m_\star$ to $m_{\star 1}$, then the collisional evolution of the transformed cluster is identical to the original one, so long as the time variable $\tau$ is replaced by the new time variable $\tau_1 = (m_{\star 1}/m_\star)\tau\,$.

\section{Resonant Relaxation}  
 
Here we provide a self--contained description of the equations governing RR, after generalizing to include orbit--averaged relativistic corrections and external perturbations; these will, of course, break the property of mass--invariance of time evolution mentioned above. 

\subsection{Collisionless limit}

The Ring mean--field Hamiltonian $H(\scrr, \tau)$ can be obtained by supplementing the Ring mean--field potential $\Phi(\scrr, \tau)$ with
general relativistic corrections and external perturbations. 
\emph{Relativistic corrections} to the gravity of the MBH cause precession of both the apse and node of a Gaussian Ring. The relativistic (secular) Hamiltonian governing the 1~PN Schwarzschild precession of the apses and the 1.5~PN Lense--Thirring precession of the apses and nodes for a spinning MBH is given by:
\begin{eqnarray}
\qquad H^{\rm rel}(I, L, L_z) &\;=\;&  H^{PN1}(I, L) \;+\; H^{PN1.5}(I, L, L_z)\,,
\nonumber\\[3ex]
\qquad H^{PN1}(I, L) &\;=\;& -\,B_1\,\frac{1}{I^3 L}\qquad\mbox{with}\qquad
B_1 \;=\; \frac{3(GM_\bullet)^4}{c^2}\frac{M_\bullet}{M}\,,
 \nonumber\\[3ex]
\qquad H^{PN1.5}(I, L, L_z) &\;=\;& B_{1.5}\,\frac{L_z}{I^3 L^3}\,,\qquad\mbox{with}\qquad
B_{1.5} \;=\; \frac{2(GM_\bullet)^5\,\chi}{c^3}\frac{M_\bullet}{M}\,,
\label{hrel}
\end{eqnarray}
where $0\leq \chi\leq 1$ is the spin parameter. Here we have assumed that MBH spin angular momentum points along the $z$--axis;  factors
$(M_\bullet/M) = \varepsilon^{-1}$ have been included in the constants $B_1$ and $B_{1.5}$, because we measure time in terms of the slow time $\tau=\varepsilon t\,$. \emph{External perturbations} due to nuclear density cusps and/or slowly moving distant masses can also be included. Let $\varepsilon\Phi^{\rm ext}(\scrr, \tau)\,$ be the orbit--averaged potential and $\varepsilon\bfA^{\rm ext}_\bullet(\tau)$ be the acceleration of the MBH. These external sources will contribute a (scaled) orbit--averaged tidal potential, 
\begin{eqnarray}
\Phi^{\rm tid}(\scrr, \tau) &\;=\;& \Phi^{\rm ext}(\scrr, \tau) \;+\; \bfX(\scrr)\cdot\bfA^{\rm ext}_\bullet(\tau)\,,
\label{tidpot}\\[1ex]
\mbox{where}\qquad\bfX(\scrr) &\;=\;& \oint \frac{{\rm d}w}{2\pi}\,\bfr(\scrd) \;\;=\;\; \mbox{centre--of--mass of a Gaussian Ring.}
\label{ring-cg}
\end{eqnarray}

Then the general mean--field Ring Hamiltonian is: 
\beq
H(\scrr, \tau) \;=\; \Phi(\scrr, \tau) \;+\; H^{\rm rel}(I, L, L_z) \;+\; \Phi^{\rm tid}(\scrr, \tau)\,,
\label{secham}
\eeq
where we note that the mean--field potential is, as given in eqn.(\ref{phislow}), determined self--consistently by the Ring DF:
\beq
\Phi(\scrr, \tau) \;=\; \int F(\scrr', \tau)\,\Psi(\scrr, \scrr')
{\rm d}\scrr'\,,\qquad\quad\mbox{Ring mean--field potential;}
\label{phislow-r}
\eeq
where
\beq
\Psi(\scrr, \scrr') \;=\; -GM_\bullet\oint\oint\frac{{\rm d}w}{2\pi}\,
\frac{{\rm d}w'}{2\pi}\,\frac{1}{\left|\bfr - \bfr'\right|}\,,
\qquad\quad\mbox{``bare'' inter--Ring potential.} 
\label{bare-ring}
\eeq
The (mean--field) Ring equations of motion are:
\begin{eqnarray}
I &\;=\;& \sqrt{GM_\bullet a} \;=\; \mbox{constant}\,,
\nonumber\\[1ex]
\frac{{\rm d}L}{{\rm d}\tau} &\;=\;& -\,\frac{\p H}{\p g}\,,\qquad\quad  
\frac{{\rm d}g}{{\rm d}\tau} \;=\; \frac{\p H}{\p L}\,;\qquad\quad
\frac{{\rm d}L_z}{{\rm d}\tau} \;=\; -\,\frac{\p H}{\p h}\,,\qquad\quad  
\frac{{\rm d}h}{{\rm d}\tau} \;=\; \frac{\p H}{\p L_z}\,.
\label{eom-ring}
\end{eqnarray}
Therefore, in the collisionless limit, the appropriate generalization of eqn.(\ref{cbe-sec}) for the $F(\scrr, \tau)$ is:  
\beq
\frac{\p F}{\p \tau} \;+\; \left[\,F\,,\,H\,\right] \;=\; 0\,.
\label{cbe-ring}
\eeq
This is the general form of the Ring CBE, whose properties are discussed in more detail in Paper~I.

\subsection{Kinetic equation for Resonant Relaxation}

We now need to generalize eqn.(\ref{eqn-gil-wake-kepW4}) for the Ring wake
function, and (\ref{bbgky-gil-kepF4})  for the Ring DF, to account for the change in the mean--field Hamiltonian from $\Phi(\scrr, \tau)$ to $H(\scrr, \tau)$. The new wake equation can be determined from the collisionless 
response of the stellar system, by extending the \emph{gedanken} experiment 
of \citet{gil68} for point mass stars to Gaussian Rings: from the $N$ Rings that are distributed in 5--dim Ring space according to the DF $F(\scrr, \tau)$, select one Ring and place it at the phase space location $\scrr'$; this will induce a small perturbation $(1/N)W$ at every location $\scrr$ in the 5--dim $\scrr$--space. Then the net perturbed DF at any instant $\tau' < \tau$ is: 
\beq
F_1(\scrr, \tau') \;=\; -\frac{F(\scrr, \tau')}{N} \;+\; 
\frac{\delta(\scrr - \scrr'(\tau'))}{N} \;+\; 
\frac{W(\scrr\,\vert\,\scrr'(\tau'), \tau')}{N}\,.
\label{pert-rg}
\eeq
where $\scrr'(\tau')$ is the location at time $\tau'$, of the Ring which 
arrives at $\scrr'$ at time $\tau$. Since $N \gg 1$ the perturbation is effectively infinitesimal, so $F_1$ must satisfy the linearized CBE discussed in Paper~I. This then gives the general Ring wake equation we
seek:
\begin{eqnarray}
&&\frac{\p W}{\p \tau'} \;+\; \left[\,W(\scrr\,\vert\,\scrr'(\tau'), \tau')\,,\,H(\scrr, \tau')\,\right]
\;+\; \left[\,F(\scrr, \tau')\,,\,\Phi^{\rm w}(\scrr, \scrr'(\tau'), \tau')\,\right]
\nonumber\\[1em]
&&\qquad\qquad \;=\; \left[\,\Phi^{\rm p}(\scrr, \scrr'(\tau'),\tau')\,,\,F(\scrr, \tau')\,\right]
\,,\qquad\quad\mbox{for $\tau'\;\leq\; \tau$,}
\nonumber\\[1em]
&&\mbox{with adiabatic turn--on initial condition}\quad\lim_{\tau'\to -\infty} W(\scrr\,\vert\,\scrr'(\tau'), \tau') \;=\; 0\,.
\label{eqn-gil-wake-kepW-rr}
\end{eqnarray}
Here $\Phi^{\rm w}$ is the gravitational potential due to the wake which
was derived in (\ref{phislow-w}):
\beq
\Phi^{\rm w}(\scrr, \scrr', \tau') \;=\; \int \,W(\scrr''\,\vert\,\scrr', \tau')\,\Psi(\scrr, \scrr'')\,{\rm d}\scrr''\,,\qquad\quad\mbox{Ring wake potential}\,.
\label{phislow-rw}
\eeq
$\Phi^{\rm p}$ is the difference between the ``bare'' inter--Ring interaction potential and the mean--field potential, which was derived in (\ref{phislow-p}):
\beq
\Phi^{\rm p}(\scrr, \scrr', \tau') \;=\; \Psi(\scrr, \scrr') \;-\; \Phi(\scrr, \tau')\,,\qquad\quad\mbox{Ring perturbing potential}\,.
\label{phislow-rp} 
\eeq
As earlier the Ring wake function satisfies the two identities
\begin{eqnarray}
\int W(\scrr\,\vert\, \scrr', \tau)\,\rmd\scrr &\;=\;& 0\,,\qquad\mbox{Zero mass in the wake of Ring $\scrr'$;}\nonumber\\
\int W(\scrr\,\vert\, \scrr', \tau)F(\scrr', \tau)\,\rmd\scrr' &\;=\;& 0\,,
\qquad\mbox{Zero net wake at $\scrr$ due to all the Rings.}
\end{eqnarray}
When we compare eqn.(\ref{eqn-gil-wake-kepW-rr}) with the earlier 
eqn.(\ref{eqn-gil-wake-kepW4}) for the Ring wake function, we see that the 
only change is in the second term on the left hand side where $\Phi$
has been replaced by $H$. This is expected because the mean--field Ring orbits are now governed by $H$. However on the right hand side, the 
perturbing potential $\Phi^{\rm p} = \Psi - \Phi$ remains unchanged, 
because this is the driver of the irreversible statistical evolution 
of the system, and cannot be affected by deterministic agencies like 
relativistic corrections and external gravitational fields.

The new kinetic equation for the DF $F(\scrr,\tau)$ is obtained by replacing, on the  left hand side of eqn.(\ref{bbgky-gil-kepF4}), the quantity $\left(\Phi - \Phi/N\right)$ by $\left(H - \Phi/N\right)$, because the $\Phi/N$ term corrects for the fact that only $(N-1)$ of the $N$ stars exert forces on any given star. The right hand side, which represents irreversible statistical evolution, remains unaltered. Therefore the 
general Ring kinetic equation is:
\begin{eqnarray}
&&\frac{\p F}{\p \tau} \;+\;
\left[\,F\,,\, H - \frac{\Phi(\scrr, \tau)}{N}\,\right] \;=\; 
\frac{1}{N}\int\left[\,\Psi(\scrr, \scrr')\,,\, F^{(2)}_{\rm irr}(\scrr, \scrr', \tau)\,\right]\,\rmd\scrr'\,,
\nonumber\\[1em] 
&&F^{(2)}_{\rm irr}(\scrr, \scrr', \tau) \;=\; 
W(\scrr\,\vert\,\scrr', \tau)\,F(\scrr', \tau) \;+\;
W(\scrr'\,\vert\,\scrr, \tau)\,F(\scrr, \tau) 
\nonumber\\[1ex]
&&\qquad\qquad\qquad\qquad\qquad\;+\; \int W(\scrr\,\vert\,\scrr'', \tau)\,W(\scrr'\,\vert\,\scrr'', \tau)\,F(\scrr'', \tau)\,\rmd\scrr''\,.
\label{bbgky-gil-kepF-rr}
\end{eqnarray}
By eliminating $F^{(2)}_{\rm irr}$ and manipulating the resulting 
expressions, we can also write the Ring kinetic equation in a form 
where the  dissipative and fluctuating contributions to the Ring 
collision term are displayed explicitly:
\begin{subequations}
\begin{eqnarray}
&&\frac{\p F}{\p \tau} \;+\; 
\left[\,F\,,\, H - \frac{\Phi(\scrr, \tau)}{N}\,\right] \;=\; 
C^{\rm dis}[F] \;+\; C^{\rm fluc}[F]\,,
\label{eqn-gke-kepF-rr}\\[1em]
&&C^{\rm dis}[F] \;=\; \frac{1}{N}\int 
\left[\,\Psi(\scrr, \scrr')\,,\,F(\scrr, \tau)W(\scrr'\,\vert\,\scrr, \tau)\,\right]\,\rmd\scrr'\,,
\label{coll-dis-kepF-rr}\\[1em]
&&C^{\rm fluc}[F] \;=\; \frac{1}{N}\int\,F(\scrr', \tau)\left[\,\Psi(\scrr, \scrr') \,+\, \Phi^{\rm w}(\scrr, \scrr', \tau)\,,\,W(\scrr\,\vert\,\scrr', \tau)\,\right]\rmd\scrr'\,. 
\label{coll-fluc-kepF-rr}
\end{eqnarray}
\end{subequations}
The kinetic equation --- either eqn.(\ref{bbgky-gil-kepF-rr}) or eqns.(\ref{eqn-gke-kepF-rr})--(\ref{coll-fluc-kepF-rr}) --- applies to the collisional evolution of $F$ whenever it satisfies
\beq
\left[\,F(\scrr, \tau)\,,\,H(\scrr, \tau) - \frac{\Phi(\scrr, \tau)}{N}\,\right] \;\simeq\; 0\,,\quad\mbox{being of order $1/N$ or smaller.}
\label{qstead-gil-rr}
\eeq
Therefore $F(\scrr, \tau)$ may be thought of as passing through a sequence of collisionless equilibria in a quasi--static manner.

\medskip
\noindent
{\bf \emph{Boundary conditions}:} Eqn.(\ref{eqn-gil-wake-kepW-rr}) and either eqn.(\ref{bbgky-gil-kepF-rr}) or eqns.(\ref{eqn-gke-kepF-rr})--(\ref{coll-fluc-kepF-rr}) need to be supplied with boundary conditions in $\scrr$--space. There are two main cases of interest:
\begin{itemize}
\item[{\bf 1.}] ``Lossless'' stellar systems in which the MBH is not considered to be a sink of stars. Then the DF is normalized as $\int F(\scrr, \tau)\,\rmd\scrr =\ 1\,$. Subject to this normalization $F(\scrr, \tau)$ can take any positive value at any location in $\scrr$--space; in other words, the domain of $F$ is all of $\scrr$--space. Hence the domain of the wake function $W(\scrr\,\vert\,\scrr', \tau)$ is all of $\scrr$ and $\scrr'$ spaces.  

\item[{\bf 2.}] Stellar systems in which stars that come too close to the MBH are either tidally shredded or swallowed whole. Since the system loses stars, we can only require that the normalization of the DF at some initial time $\tau_0$, be
\beq 
\int F(\scrr, \tau_0)\,\rmd\scrr \;=\; 1\,,\qquad
\mbox{Initial normalization with loss of stars to MBH.}
\label{F-norm}
\eeq 
At later times $\tau > \tau_0$, we will have $\int F(\scrr, \tau)\,\rmd\scrr \;\leq\; 1\,$. The simplest model of the loss is the assumption that a star is lost to the MBH when its pericentre distance is smaller than some fixed value $r_{\rm lc}$, the \emph{loss--cone radius}. When the loss--cone is empty, stars belonging to the cluster must necessarily have pericentre radii $a(1-e)$ larger than $r_{\rm lc}$. Hence the domain of $F(\scrr, \tau)$ is restricted to regions of $\scrr$--space in which $I$ and $L$ are large enough: 
\begin{eqnarray}
\qquad I_{\rm lc} \;<\; I &\qquad\mbox{where}\qquad& I_{\rm lc} \;=\; \sqrt{GM_\bullet r_{\rm lc}}\;;\nonumber\\[1ex]  
\qquad L_{\rm lc}(I) \;<\; L \;\leq\; I\,, &\qquad\mbox{where}\qquad&
L_{\rm lc}(I) \;=\; I_{\rm lc}\left[\,2 \,-\, \left(\frac{I_{\rm lc}}{I}\right)^2\,\right]^{1/2}\,.
\end{eqnarray}
As $I$ increases from its minimum value of $I_{\rm lc}$, the function 
$L_{\rm lc}(I)$ monotonically increases from its minimum value of 
$I_{\rm lc}$ and approaches its asymptotic value of $\sqrt{2}I_{\rm lc}\,$.
The Ring DF and wake functions satisfy the ``empty loss--cone'' (or ``absorbing'') boundary conditions: 
that for all $\tau$,  
\begin{eqnarray}
\qquad F(\scrr, \tau) &\;=\;& 0\qquad\mbox{for $I \leq I_{\rm lc}\,$ and $\,L \leq L_{\rm lc}(I)$,}\nonumber\\[1ex]
\qquad W(\scrr\,\vert\,\scrr', \tau) &\;=\;& 0\qquad\mbox{for $I\,, I' \leq I_{\rm lc}\,$ and $\,L \leq L_{\rm lc}(I)$, $\,L' \leq L_{\rm lc}(I')$.}
\label{elc-bc}
\end{eqnarray}
\end{itemize} 

Eqn.(\ref{eqn-gil-wake-kepW-rr}) and either eqn.(\ref{bbgky-gil-kepF-rr}) or (\ref{eqn-gke-kepF-rr})--(\ref{coll-fluc-kepF-rr}), together with suitable boundary conditions on the DF and wake, are the fundamental equations governing Resonant Relaxation.

\section{Discussion}

It should come as no surprise that there is a structural similarity between the Ring kinetic equation, and the Poisson Bracket form of Gilbert's equation.\footnote{Compare the Ring wake eqn.(\ref{eqn-gil-wake-kepW-rr}) with eqn.(\ref{eqn-gil-wake-pb}); the Ring kinetic equation in the BBGKY form eqn.(\ref{bbgky-gil-kepF-rr}) with eqn.(\ref{bbgky-gil}); or the Ring kinetic equation in fluctuation--dissipation form eqns.(\ref{eqn-gke-kepF-rr})--(\ref{coll-fluc-kepF-rr}) with eqns.(\ref{eqn-gke-pb})--(\ref{coll-fluc-pb}).} This invariance of form is a natural consequence of averaging over a (fast) angle variable, a procedure that preserves canonical structure in phase space. Gilbert's equations apply to a general stellar system in 6--dim phase space, whereas the RR equations apply to a Keplerian stellar system orbiting a MBH in a reduced 5--dim Ring space. What we have demonstrated is that the Ring equations can be got from the 6--dim Gilbert equations by  replacing (a) ``point mass star'' with ``Gaussian Ring'', (b) all 6--dim PBs with 4--dim PBs, and (c) the time variable $t$ by the slow time $\tau=\varepsilon t$. Thus we have corresponding (scaled) potentials: $\Psi(\scrr, \scrr')$ which is the ``bare'' inter--Ring potential between Rings $\scrr$ and $\scrr'$; the Ring mean--field potential $\,\Phi(\scrr, \tau)$ at location $\scrr$; the potential $\Phi^{\rm w}(\scrr, \scrr', \tau)$ felt by a Ring at $\scrr$ due to the wake of Ring $\scrr'$; and $\Phi^{\rm p}(\scrr, \scrr', \tau) = \Psi(\scrr, \scrr') - \Phi(\scrr, \tau)$ which is the perturbing potential felt by a Ring at $\scrr$ due to a Ring at $\scrr'$. All these potentials are related to the DF or the wake through formulae that are analogous to those in Gilbert's theory.

\medskip
\noindent
{\bf \emph{Ring Wake as the driver of RR}:} The kinetic equation in the BBGKY form eqn.(\ref{bbgky-gil-kepF-rr}) tells us that irreversible collisional evolution is  driven by the irreducible part of the 2--particle correlation: 
\begin{eqnarray}
F^{(2)}_{\rm irr}(\scrr, \scrr', \tau) &\;=\;& 
W(\scrr\,\vert\,\scrr', \tau)\,F(\scrr', \tau) \;+\;
W(\scrr'\,\vert\,\scrr, \tau)\,F(\scrr, \tau) 
\nonumber\\[1ex]
&&\qquad\qquad\;+\; \int W(\scrr\,\vert\,\scrr'', \tau)\,W(\scrr'\,\vert\,\scrr'', \tau)\,F(\scrr'', \tau)\,\rmd\scrr''\,,
\label{irr-kep0-discussion}
\end{eqnarray}
This decomposition of $F^{(2)}_{\rm irr}$ means that the irreducible 2--Ring correlation at $(\scrr, \scrr')$ gets three kinds of contributions from the wake function: (a) The wake of $\scrr'$ at $\scrr$; (b) The wake of $\scrr$ at $\scrr'$; (c) The product of the wake values at the points $\scrr$ and $\scrr'$ of a third Ring at $\scrr''$, summed over all locations $\scrr''$. All three contributions come with suitable $F$--weighting. The third term accounts for the contribution of collective effects (``gravitational polarization'') to the microscopic processes driving RR. This is the nature of the full theory at $O(1/N)$.

Correlations build through Ring--Ring collisions: very early times $\tau\to -\infty$, the wake $W \to 0$ which implies that $F^{(2)}_{\rm irr} \to 0$. As the wake of every Ring builds over time, so does $F^{(2)}_{\rm irr}$, so we can think of the wake as the fundamental driver of collisional evolution. The Ring wake eqn.(\ref{eqn-gil-wake-kepW-rr}) is:
\begin{eqnarray}
&&\frac{\p W}{\p \tau'} \;+\; \left[\,W(\scrr\,\vert\,\scrr'(\tau'), \tau')\,,\,H(\scrr, \tau')\,\right]
\;+\; \left[\,F(\scrr, \tau')\,,\,\Phi^{\rm w}(\scrr, \scrr'(\tau'), \tau')\,\right]
\nonumber\\[1em]
&&\qquad\qquad \;=\; \left[\,\Phi^{\rm p}(\scrr, \scrr'(\tau'),\tau')\,,\,F(\scrr, \tau')\,\right]
\,,\qquad\quad\mbox{for $\tau'\;\leq\; \tau$,}
\nonumber\\[1em]
&&\mbox{with adiabatic turn--on initial condition}\quad\lim_{\tau'\to -\infty} W(\scrr\,\vert\,\scrr'(\tau'), \tau') \;=\; 0\,,
\nonumber
\end{eqnarray}
where 
\begin{eqnarray}
\Phi^{\rm w}(\scrr, \scrr', \tau') &\;=\;& \int \,W(\scrr''\,\vert\,\scrr', \tau')\,\Psi(\scrr, \scrr'')\,{\rm d}\scrr''\,,\qquad\quad\mbox{Ring wake potential;}
\nonumber\\[1ex]
\Phi^{\rm p}(\scrr, \scrr', \tau') &\;=\;& \Psi(\scrr, \scrr') \;-\; \Phi(\scrr, \tau')\,,\qquad\quad\mbox{Ring perturbing potential.}
\nonumber 
\end{eqnarray}
Note that $\,\Phi^{\rm w}\,$ is linear in $W$, whereas $\,\Phi^{\rm p}\,$ is independent of $W$. If the right hand side happened to be zero, i.e.
$\left[\Phi^{\rm p}, F(\scrr, \tau')\right] = 0$, then the wake equation 
would be a linear integral equation which is homogeneous in $W$. The solution that is compatible with the adiabatic turn--on initial condition 
is $W=0\,$ for all time. Since there is no wake, the collision term 
in eqn.(\ref{eqn-gke-kepF-rr})--(\ref{coll-fluc-kepF-rr}) vanishes. 
Then the Ring DF satisfies the Ring CBE, and there is no relaxation. 
Therefore 
\beq
\left[\,\Phi^{\rm p}(\scrr, \scrr'(\tau'),\tau')\,,\,F(\scrr, \tau')\,\right] \;=\;
\left[\,\Psi(\scrr, \scrr'(\tau')) \,-\, \Phi(\scrr, \tau')\,,\,F(\scrr, \tau')\,\right] 
\label{source}
\eeq
is the ``source term'' for the Ring wake function, and for the entire RR process. In general this drives changes in both the magnitude and direction 
of angular momenta through apsidal and nodal resonances, and there is 
no strict separation into scalar--RR and vector--RR.

\medskip
\noindent
{\bf \emph{Time evolution}:} The Ring kinetic equation is for RR what the 
Boltzmann equation is to the kinetic theory of gases. Just as the Boltzmann
equation follows the DF over times that are longer than the duration of collisions between gas molecules, the Ring kinetic equation tracks changes in the DF of the system over times that are longer than $\ts$ which is the typical coherence time of secular stellar encounters. Over times $\sim \mbox{several}\; \ts \ll N\ts$, the collision term $\left\{C^{\rm dis}[F] + C^{\rm fluc}[F]\right\}$ makes only a small contribution to the change in $F$. Then the system is effectively collisionless and is well--approximated by the Ring CBE eqn.(\ref{cbe-ring}). In this phase the Ring DF can display significant variations over times $\ts$: these could be due to (a) secular instabilities, or (b) secular collisionless relaxation (i.e. violent relaxation of Gaussian Rings), or (c) excitation of secular modes by an external perturber, and possible loss of stars to the MBH. In the absence of continuing forcing on the secular time scale by external sources, it is expected that the stellar system would settle in a stationary state $\p F/\p \tau = 0\,$ which implies that $[F, H] = 0$. By the secular Jeans theorem of Paper~I, $F$ must be a function of the isolating integrals of motion of $H$. These Ring DFs are necessarily linearly stable to perturbations. 

Over a longer time scale $\gg \ts$ the effect of the collision term 
$\left\{C^{\rm dis}[F] + C^{\rm fluc}[F]\right\}$ can no longer be ignored. 
The Ring DF will evolve slowly due to the slow accrual of the relaxing effect of angular momentum exchanges between the Rings. We can think of $F$ as passing slowly through a sequence of secular Jeans equilibria for which 
$\left[F\,,\, H - \Phi/N\right]$ is of smaller order than $1/N$. 
Then eqns.(\ref{eqn-gke-kepF-rr})--(\ref{coll-fluc-kepF-rr}) imply that 
$\p F/\p\tau \,\simeq\, \left\{C^{\rm dis}[F] + C^{\rm fluc}[F]\right\} \,\sim\, O(1/N)\,$. Therefore the time over which $F$ deforms (quasi--steadily) by order unity is the RR time scale $T_{\rm res} \,=\, N\ts \,=\, (N/\varepsilon)\tk\,$. During this phase of evolution we can allow for the external tidal potential $\Phi^{\rm tid}$ to also vary slowly --- as might be the case when a galactic nuclear cusp adjusts adiabatically to the slowly varying Ring DF. All through the relaxation process, Rings exchange angular momentum, so a Ring that has lost enough angular momentum to have its periapse decrease to $r_{\rm lc}$ will be lost to the MBH. It is of great interest to determine the mass, energy and angular momentum lost by the Keplerian star cluster to the MBH.

\section{Conclusions}

We have derived the fundamental kinetic equation governing the Resonant 
Relaxation (RR) of low mass stellar systems around a massive black hole (MBH). This includes the effects of stellar self--gravity, general relativistic corrections up to 1.5 post--Newtonian (PN) order, as well as external deterministic sources of gravity that vary on the secular time scale. We considered a stellar system consisting of stars of equal mass, but this is not a serious limitation. Mass--segregation effects in RR can be studied by straightforward generalization to a system with a range of stellar masses. We have seen in the discussion of the previous section that 
the wake function is the driver of RR, and that the source term for the wake has contributions from both apsidal and nodal resonances which may be equally important in general geometries. Therefore the traditional, physically--motivated split of RR into scalar--RR and vector--RR dissolves, in general, into a seamless blend of both processes, with the later dominating when fast apse precession promotes apse--averaged dynamics. We noted the structural similarity between the Ring kinetic equation and Gilbert's equation. \citet{hey10} proved an H--theorem for the latter, for an integrable stellar system admitting global action--angle variables. It seems reasonable to expect that a similar exercise would be successful for RR. Our theory of RR is valid for stellar systems with arbitrary geometry and figure rotation, so long as these are dynamically stable quasi--steady secular equilibria. The orbital structure can be regular, chaotic, or mixed;\footnote{Averaging over the fast Kepler orbital phase is independent of whether the resulting secular dynamics is integrable or not; this is as true for planetary systems as Keplerian stellar systems. When the secular dynamics is non--integrable, chaotic diffusion through broken tori is
restricted to the 4--dim $\{L, L_z, g, h\}$ subspace of the full 6--dim 
phase space.} the formalism presented in this paper applies to all of these. 

Below is a brief guide for the application of the Ring kinetic equation to any Keplerian stellar system. These are followed by remarks on future 
directions and useful ways of comparison with numerical simulations.
\begin{itemize}
\item[{\bf 1.}] Secure an ``explicit'' form for the orbit--averaged Poisson kernel $\Psi$ in eqn.~(\ref{bare-ring}). Although desirable, a closed form expression is difficult to come by; more often than not the orbit averaged ``bare'' Ring potential is expressed as a Fourier series in the apses and nodes, with the Fourier coefficients depending on the eccentricities and the inclinations (or related Delaunay variables). One could also work with a pre--computed table of the kernel in the physical limit of interest.
 
\item[{\bf 2.}] Using $\Psi$ construct an equilibrium pair, $\{F\,, \Phi\}\,$, of DF $F$ and mean--field potential, $\Phi$, which is related to $F$
through the integral of eqn.(\ref{phislow-r}). Secular dynamics is governed by $H$, the mean--field Hamiltonian of eqn.(\ref{secham}), which is the sum of the mean--field potential $\Phi$, and contributions from 1.5~PN general relativistic corrections and any external gravitational potential.
By the secular Jeans theorem of Paper~I, $F$ must be a function of $H$. In general one needs to solve an integral equation; it would be very interesting to see the development of Schwarzschild--like iterative numerical methods. Moreover it is also necessary that the $F(H)$ be dynamically stable.

\item[{\bf 3.}] With orbit--averaged kernel in hand, and an initially stable equilibrium model, solve the wake equation (\ref{eqn-gil-wake-kepW-rr}) for $W$, with adiabatic turn on initial condition.

\item[{\bf 4.}] Use the pair $\{F\,, W\}$ to compute the collision integral, 
which is the right hand side of the Ring kinetic equation (\ref{eqn-gke-kepF-rr})--(\ref{coll-fluc-kepF-rr}) driving relaxation.

\item[{\bf 5.}] Solve the kinetic equation (Eq.~\ref{eqn-gke-kepF-rr})--(\ref{coll-fluc-kepF-rr}) for $F\,$: the global solution of eqn.(\ref{eqn-gke-kepF-rr}) couples all the three steps given above, with the Ring 
mean--field potential $\Phi$ recovered from $F$, then wake $W$ updated with $F$ and $\Phi$ as one progresses towards a relaxed state.\footnote{In the course of relaxation, the linear secular stability of the evolving quasi--equilibria must be monitored. In case one encounters a dynamical instability, fast (i.e. on the secular time scale) collisionless processes will intervene. Once the instability has saturated and the system settled in a new secular collisionless equilibrium, the subsequent RR evolution can be resumed by following steps 2-5.}
\end{itemize}
The steps above are explicitly laid out in a companion paper \citet{st16},
where we formulate and study the physical kinetics of the RR of zero--thickness, flat, axisymmetric discs. Working with the \citet{ps82} approximation of neglecting ``gravitational polarization'' we derive a 
Fokker--Planck equation for the DF with  diffusion coefficients that are given self--consistently in terms of contributions from apsidal resonances between pairs of stellar orbits. For `lossless' stellar discs we 
prove an H--theorem and conservation laws for the disc mass, energy and angular momentum. When stars are lost to the MBH through an empty loss--cone, the loss rates of  mass, energy and angular momentum can be expressed in terms of the diffusion flux at the loss--cone boundary.

It is hard to overemphasize the need for comparison with numerical simulations, and we can see three different routes of exploration: (a) The diffusion coefficients and angular momentum diffusion statistics predicted by our RR theory could be compared with results obtained from refined N--body simulations designed for that same purpose, such as \citet{mamw11} for RR in a spherical cluster; (b) The Ring kinetic equation (\ref{eqn-gke-kepF-rr})--(\ref{coll-fluc-kepF-rr}) can be solved for the thermally relaxed equilibrium of any chosen initial stable equilibrium. The equilibrium and associated statistics could be compared with the N--body evolution of (a sample of) the same initial state; (c) The time evolving DF of our formalism can be compared with related solutions of various Fokker--Planck equations proposed by others \citep{ha06, mhl11}, \emph{ad hoc} though their assumptions may be. This should allow calibration of the performance of \emph{ad hoc} recipes for RR vis--a--vis the fundamental formulation 
we have presented. 

\section*{Acknowledgments}
We are grateful to Jerome Perez, Stephane Colombi and the Institut Henri Poincar\'e for hosting us when a part of this work was done. We thank Scott Tremaine for comments on an earlier draft.

\appendix
\section{Derivation of the orbit--averaged kinetic equation}

Here we derive the orbit--averaged kinetic equation, beginning with the exact Gilbert equations eqns.(\ref{eqn-gil-wake-del}) and (\ref{bbgky-gil-del}) given in Delaunay variables. Orbit--averaging is a systematic procedure, of perturbative development of these equations in the small parameter $\varepsilon$. 

The first step is to substitute in (\ref{bbgky-gil-del}) the  multiple--scale forms, eqn.(\ref{dffou}) for the DF $f$ and  eqn.(\ref{wake-gr}) for the wake $\scrw$. Then all the $O(1)$ terms vanish. The $O(\varepsilon)$ terms give the following equation for $F(\scrr, \tau)$ and all the $f_n(\scrr, \tau)$:
\begin{eqnarray}
&&\frac{\p F}{\p \tau} \;+\; \sum_{n \neq 0}{\rm i}n\Omega_{\rm k}\,f_n\exp{[{\rm i}n w]} \;+\;
\left[\,F\,,\,\left(1 - \frac{1}{N}\right)\varphi \,+\, \bfr\cendot\bfA_\bullet\,\right]_{(6)}\;=\; \nonumber\\[1ex]
&&\qquad\qquad\qquad\qquad\qquad\frac{1}{N}\int \left[\,p(\scrd, \scrd')\,,\, F^{(2)}_{\rm irr}(\scrr, \scrr', \tau)\,\right]_{(6)}\,\rmd\scrr'\,\frac{\rmd w'}{2\pi}\,,
\label{bbgky-gil-kep0}
\end{eqnarray}
where 
\begin{eqnarray}
F^{(2)}_{\rm irr}(\scrr, \scrr', \tau) &\;=\;& 
W(\scrr\,\vert\,\scrr', \tau)\,F(\scrr', \tau) \;+\;
W(\scrr'\,\vert\,\scrr, \tau)\,F(\scrr, \tau) 
\nonumber\\[1ex]
&&\qquad\qquad\;+\; \int W(\scrr\,\vert\,\scrr'', \tau)\,W(\scrr'\,\vert\,\scrr'', \tau)\,F(\scrr'', \tau)\,\rmd\scrr''\,,
\label{irr-kep0}
\end{eqnarray}
is the irreducible part of the 2--Ring correlation function. Note that only the $O(1)$ part of the wake $W$ contributes directly to the $O(\varepsilon)$ kinetic theory. The Poisson kernel $p(\scrd, \scrd')= -GM_\bullet/\vert\bfr - \bfr'\vert\,$ is a specified $O(1)$ function of its arguments and is independent of $\varepsilon$. The other quantities, $\varphi$ and $\bfA_\bullet$, are needed only to $O(1)$ accuracy. From eqn.(\ref{accn-del}) we have to $O(1)$:
\beq
\bfA_\bullet(\tau) \;=\; GM_\bullet\int
F(\scrr, \tau)\,{\rm d}\scrr
\oint \frac{{\rm d}w}{2\pi}\,\frac{\hat{\bfr}\;}{r^2}
\;=\; {\bf 0}\,,
\label{azero}
\eeq
because $\oint {\rm d}w\,\hat{\bfr}/r^2 = {\bf 0}\,$, by the conservation of angular momentum along a Kepler orbit. Recalling that $\ddot\bfx = \varepsilon\bfA_\bullet$ we see that, as in the collisionless 
theory of Paper~I, the MBH does not accelerate to $O(\varepsilon)$ accuracy. From eqn.(\ref{mean-pot-del}) the rescaled mean potential 
to $O(1)$ is:
\beq
\varphi(\scrd, \tau) \;=\; \int F(\scrr', \tau)\,\rmd\scrr'\oint
p(\scrd, \scrd')\,\frac{\rmd w'}{2\pi}\,.
\label{mean-pot-kep}
\eeq
The PB on the right hand side of eqn.(\ref{bbgky-gil-kep0}) are over the 
$\scrd$ variables, so the integral over $w'$ can be taken inside the PB. Since $p(\scrd, \scrd')$ is the only quantity inside the PB that depends on $w'$, the $w'$--integral operates only on $p(\scrd, \scrd')$. Then eqn.(\ref{bbgky-gil-kep0}) simplifies to:
\begin{eqnarray}
&&\frac{\p F}{\p \tau} \;+\; \sum_{n \neq 0}{\rm i}n\Omega_{\rm k}\,f_n\exp{[{\rm i}n w]} \;+\;
\left(1 - \frac{1}{N}\right)\left[\,F\,,\,\varphi(\scrd, \tau)\,\right]_{(6)}\;=\; 
\nonumber\\[1ex]
&&\qquad\qquad\qquad\qquad\qquad\frac{1}{N}\int\left[\,\oint p(\scrd, \scrd')\frac{\rmd w'}{2\pi}\,,\, F^{(2)}_{\rm irr}(\scrr, \scrr', \tau)\,\right]_{(6)}\,\rmd\scrr'\,\,.
\label{bbgky-gil-kep1}
\end{eqnarray}
Averaging over $w$, we get the following equation for $F(\scrr, \tau)$:
\begin{eqnarray}
&&\frac{\p F}{\p \tau} \;+\; 
\left(1 - \frac{1}{N}\right)\oint\left[\,F\,,\,\varphi(\scrd, \tau)\,\right]_{(6)}\frac{\rmd w}{2\pi} \;=\; 
\nonumber\\[1ex]
&&\qquad\qquad\qquad\qquad\qquad\frac{1}{N}\int\left[\,\oint\oint p(\scrd, \scrd')\frac{\rmd w}{2\pi}\frac{\rmd w'}{2\pi}\,,\, F^{(2)}_{\rm irr}(\scrr, \scrr', \tau)\,\right]_{(6)}\,\rmd\scrr'\,\,.
\label{bbgky-gil-kepF}
\end{eqnarray}
Subtracting eqn.(\ref{bbgky-gil-kepF}) from (\ref{bbgky-gil-kep1}), and 
dropping the extremely small terms proportional to $1/N$ in comparison 
to $O(1)$ terms, we get the following equation for the $f_n(\scrr, \tau)$:
\beq
\sum_{n \neq 0}{\rm i}n\Omega_{\rm k}\,f_n\exp{[{\rm i}n w]} 
\;+\; \left[\,F\,,\,\varphi(\scrd, \tau)\,\right]_{(6)} \;-\; \oint\left[\,F\,,\,\varphi(\scrd, \tau)\,\right]_{(6)}\frac{\rmd w}{2\pi} \;=\; 0\,,
\label{eqn-gke-kepfn0}
\eeq
with solution
\beq
f_n(\scrr, \tau) \;=\;  \frac{{\rm i}}{n \Omega_{\rm k}} \oint\left[\,F\,,\,\varphi(\scrd, \tau)\,\right]_{(6)}\exp{[-{\rm i}n w]}\,\frac{\rmd w}{2\pi}\,,\qquad\qquad n\neq 0\,.
\label{eqn-gke-kepfn}
\eeq
Hence all the $f_n(\scrr, \tau)$ for $n\neq 0$ are slaved to the quasi--steady collisional evolution of $F(\scrr, \tau)$. As in Paper~I, the function $f_0(\scrr, \tau)$ remains undetermined in our $O(\varepsilon)$ calculations. 

Similarly, when the expansions of eqns.(\ref{dffou}) and (\ref{wake-gr}) for the DF and the wake are substituted in eqn.(\ref{eqn-gil-wake-del}), all the $O(1)$ terms vanish. The $O(\varepsilon)$ terms give the following equation for $W(\scrr\,\vert\,\scrr'(\tau'), \tau')$ and all the $\varpi_n(\scrr\,\vert\,\scrd'(\tau'), \tau')$:
\begin{eqnarray}
&&\frac{\p W}{\p \tau'} \;+\; \sum_{n \neq 0}{\rm i}n\Omega_{\rm k}\,\varpi_n\exp{[{\rm i}n w]} \;+\; \left[\,W\,,\,\varphi(\scrd, \tau')\,\right]_{(6)} 
\;+\; \left[\,F(\scrr, \tau')\,,\,\varphi^{\rm w}(\scrd, \scrr'(\tau'), \tau')\,\right]_{(6)}\nonumber\\[1em]
&&\qquad\qquad \;=\;  
\left[\,\varphi^{\rm p}(\scrd, \scrd'(\tau'), \tau')\,,\,F(\scrr, \tau')\,\right]_{(6)}\,,\qquad\quad\mbox{for $\tau'\;\leq\; \tau\,$.}
\label{eqn-gil-wake-kep0}
\end{eqnarray}
Note that we have written the rescaled wake potential as $\varphi^{\rm w}(\scrd, \scrr', \tau)$, instead of $\varphi^{\rm w}(\scrd, \scrd', \tau)$. 
This is because, when the expansion of eqn.(\ref{wake-gr}) is substituted 
in eqn.(\ref{wake-pot-del}), we find that $\varphi^{\rm w}$ is independent of $w'$ to $O(1)$ accuracy:
\beq
\varphi^{\rm w}(\scrd, \scrr', \tau) \;=\; \int W(\scrr''\,\vert\,\scrr', \tau)\,\rmd\scrr''\oint p(\scrd, \scrd'')\,\frac{\rmd w''}{2\pi}\,.
\label{wake-pot-kep}
\eeq
In eqn.(\ref{eqn-gil-wake-kep0}) $\scrd'(\tau')=\{\scrr'(\tau'), w'(\tau')\}$ is the location of the star in 6--dim phase space at time $\tau'$, which arrives at the phase space location $\scrd'=\{\scrr', w'\}$ at time $\tau$. In general this orbit is governed by the Hamiltonian $H_{\rm org}$ of eqn.(\ref{ham0-del}), but we need it in eqn.(\ref{eqn-gil-wake-kep0}) only to $O(1)$ accuracy. This is just the secular collisionless dynamics described above by eqns.(\ref{ring-eom-sg}) and (\ref{orb-rate-sg}): $\scrr'(\tau')$ is such that $I'=\mbox{constant}\;$ and  $\{L'(\tau'), L_z'(\tau'); g'(\tau'), h'(\tau')\}$ obey the Hamiltonian equations of motion eqn.(\ref{ring-eom-sg}) with $\Phi(\scrr', \tau')$ acting as the Hamiltonian. Integrating eqn.(\ref{orb-rate-sg}) over time, we have 
\beq
w'(\tau') \;=\; w' \;+\; \frac{\Omega_{\rm k}(I')}{\varepsilon}\left(\tau' - \tau\right) \;+\; \int_{\tau}^{\tau'}\frac{\p \Phi(\scrr'(\tau''), \tau'')}{\p I}\,\rmd\tau''\,. 
\label{orb-phas-sg}
\eeq
The Ring orbit $\scrr'(\tau')$ is independent of $w'(\tau')$, which is 
slaved to it.

Since only the $W$ part of the wake contributes to the equation for $F$ and $f_n$, we can obtain an equation for it by averaging eqn.(\ref{eqn-gil-wake-kep0}) over both $w$ and $w'$. All the terms with $\varpi_n$ vanish in the $w$--averaging, and play no further role. Then the equation for $W(\scrr\,\vert\,\scrr'(\tau'), \tau')$ is:
\begin{eqnarray}
&&\frac{\p W}{\p \tau'} \;+\; \oint\!\!\oint\left[\,W\,,\,\varphi(\scrd, \tau')\,\right]_{(6)}\,\frac{\rmd w}{2\pi}\frac{\rmd w'}{2\pi} 
\;+\; \oint\!\!\oint\left[\,F(\scrr, \tau')\,,\,\varphi^{\rm w}(\scrd, \scrr'(\tau'), \tau')\,\right]_{(6)}\,\frac{\rmd w}{2\pi}\frac{\rmd w'}{2\pi}\nonumber\\[1em]
&&\qquad\qquad \;=\;  
\oint\!\!\oint\left[\,\varphi^{\rm p}(\scrd, \scrd'(\tau'), \tau')\,,\,F(\scrr, \tau')\,\right]_{(6)}\,\frac{\rmd w}{2\pi}\frac{\rmd w'}{2\pi}\,,\qquad\quad\mbox{for $\tau'\;\leq\; \tau$.}
\nonumber\\[1em]
&&\mbox{with adiabatic turn--on initial condition}\quad\lim_{\tau'\to -\infty} W(\scrr\,\vert\,\scrr'(\tau'), \tau') \;=\; 0\,.
\label{eqn-gil-wake-kepW}
\end{eqnarray}

Equation (\ref{eqn-gil-wake-kepW}) for $W$, eqn.(\ref{bbgky-gil-kepF}) for $F$, eqn.(\ref{irr-kep0}) for $F^{(2)}_{\rm irr}$ and eqn.(\ref{eqn-gke-kepfn}) for the $f_n$ are the orbit--averaged equations describing the $O(\varepsilon)$ kinetic theory. However they are in a raw form and need to be processed further such that all direct dependences on the orbital phases $w$ and $w'$ are eliminated. In other words, we seek a kinetic description that is self--consistent and closed in 5--dim $\scrr$--space. This is done in below by (a) expressing the 6--dim PBs in terms of 4--dim PBs, and (b) defining suitable potential functions in $\scrr$--space.

For any two phase space functions, $\chi_1(\scrd)$ and $\chi_2(\scrd)$, the 6--dim PB can be expressed in terms of the 4--dim PB (without subscript) as
\beq
\left[\,\chi_1\,,\,\chi_2\,\right]_{(6)} \;=\;
\left(\frac{\p \chi_1}{\p w}\frac{\p \chi_2}{\p I} -
\frac{\p \chi_1}{\p I}\frac{\p \chi_2}{\p w}\right) \;+\;
\left[\,\chi_1\,,\,\chi_2\,\right]\,.
\label{pb624}
\eeq
For functions $\chi_1(\scrr)$ and $\chi_2(\scrr)\,$, the 6--dim PB equals the 4--dim PB: $\left[\,\chi_1\,,\,\chi_2\,\right]_{(6)} = \left[\,\chi_1\,,\,\chi_2\,\right]$.

We begin with the 3 PBs in the wake eqn.(\ref{eqn-gil-wake-kepW}):
\begin{eqnarray}
\left[\,W(\scrr\,\vert\,\scrr'(\tau'), \tau')\,,\,\varphi(\scrd, \tau')\,\right]_{(6)} &\;=\;& - \frac{\p W}{\p I}\frac{\p \varphi}{\p w} \;+\;
\left[\,W(\scrr\,\vert\,\scrr'(\tau'), \tau')\,,\,\varphi(\scrd, \tau')\,\right]\,,
\nonumber\\[1em]
\left[\,F(\scrr, \tau')\,,\,\varphi^{\rm w}(\scrd, \scrr'(\tau'), \tau')\,\right]_{(6)} &\;=\;& - \frac{\p F}{\p I}\frac{\p \varphi^{\rm w}}{\p w} \;+\; \left[\,F(\scrr, \tau')\,,\,\varphi^{\rm w}(\scrd, \scrr'(\tau'), \tau')\,\right]\,,
\nonumber\\[1em]
\left[\,\varphi^{\rm p}(\scrd, \scrd'(\tau'), \tau')\,,\,F(\scrr, \tau')\,\right]_{(6)} &\;=\;& + \frac{\p \varphi^{\rm p}}{\p w}\frac{\p W}{\p I} \;+\; \left[\,\varphi^{\rm p}(\scrd, \scrd'(\tau'), \tau')\,,\,F(\scrr, \tau')\,\right]\,,\nonumber
\end{eqnarray}
The first term on the right hand side of all three equations can be written
in the form $\p\{\,\}/\p w$, and will vanish when averaged over $w$. So 
the three double--integrals over the 6--dim PBs in eqn.(\ref{eqn-gil-wake-kepW}) reduce to double--integrals over the 4--dim PBs. Since the 4--dim PBs act only in $\scrr$--space, the integrals over $w$ and $w'$ can be taken inside the 4--dim PBs. Of the two functions in each of the PBs, it is only the potentials $\varphi$, $\varphi^{\rm w}$ and $\varphi^{\rm p}$ that 
depend on the orbital phases $w$ and $w'$. In particular $\varphi$ and $\varphi^{\rm w}$ are functions of $w$ but independent of $w'$, so 
we get:
\begin{eqnarray}
\oint\!\!\oint\left[\,W(\scrr\,\vert\,\scrr'(\tau'), \tau')\,,\,\varphi(\scrd, \tau')\,\right]_{(6)}\,\frac{\rmd w}{2\pi}\frac{\rmd w'}{2\pi} &\;=\;&
\left[\,W(\scrr\,\vert\,\scrr'(\tau'), \tau')\,,\,\oint\varphi(\scrd, \tau')\frac{\rmd w}{2\pi}\,\right]\,,\nonumber\\[1em]
\oint\!\!\oint\left[\,F(\scrr, \tau')\,,\,\varphi^{\rm w}(\scrd, \scrr'(\tau'), \tau')\,\right]_{(6)}\,\frac{\rmd w}{2\pi}\frac{\rmd w'}{2\pi} &\;=\;&
\left[\,F\,,\,\oint\varphi^{\rm w}(\scrd, \scrr'(\tau'), \tau')\frac{\rmd w}{2\pi}\,\right]\,.
\nonumber
\end{eqnarray}
However, $\varphi^{\rm p}$ depends on both $w$ and $w'$, so we are 
left with a double integral inside the 4--dim PB:
\beq
\oint\!\!\oint\left[\,\varphi^{\rm p}(\scrd, \scrd'(\tau'), \tau')\,,\,F(\scrr, \tau')\,\right]_{(6)}\,\frac{\rmd w}{2\pi}\frac{\rmd w'}{2\pi}  \;=\; \left[\,\oint\!\!\oint\varphi^{\rm p}(\scrd, \scrd'(\tau'), \tau')\frac{\rmd w}{2\pi}\frac{\rmd w'}{2\pi}\,,\,F(\scrr, \tau')\,\right]\,.
\nonumber
\eeq
In the integral over $w'$ on the right hand side, we note that $w'$ occurs in the integrand only through the function $w'(\tau')$ which is given in eqn.(\ref{orb-phas-sg}). Since $w'$ appears here only as a linear additive term, we can 
replace $w'(\tau')$ by $w'$ in the potential $\varphi^{\rm p}$: 
\beq
\oint\!\!\oint\left[\,\varphi^{\rm p}(\scrd, \scrd'(\tau'), \tau')\,,\,F(\scrr, \tau')\,\right]_{(6)}\,\frac{\rmd w}{2\pi}\frac{\rmd w'}{2\pi}  \;=\; \left[\,\oint\!\!\oint\varphi^{\rm p}(\scrd, \scrr'(\tau'), w', \tau')\frac{\rmd w}{2\pi}\frac{\rmd w'}{2\pi}\,,\,F(\scrr, \tau')\,\right]\,, 
\nonumber
\eeq
where the dependence of $\varphi^{\rm p}$ on the arguments $\scrr'(\tau')$ and $w'$ is now indicated explicitly. 

\nin
Then eqn.(\ref{eqn-gil-wake-kepW}) for $W(\scrr\,\vert\,\scrr'(\tau'), \tau')$ can be written as:
\begin{eqnarray}
&&\frac{\p W}{\p \tau'} \;+\; \left[\,W(\scrr\,\vert\,\scrr'(\tau'), \tau')\,,\,\oint\varphi(\scrd, \tau')\frac{\rmd w}{2\pi}\,\right]
\;+\; \left[\,F(\scrr, \tau')\,,\,\oint\varphi^{\rm w}(\scrd, \scrr'(\tau'), \tau')\frac{\rmd w}{2\pi}\,\right]
\nonumber\\[1em]
&&\qquad\qquad \;=\; \left[\,\oint\!\!\oint\varphi^{\rm p}(\scrd, \scrr'(\tau'), w',\tau')\frac{\rmd w}{2\pi}\frac{\rmd w'}{2\pi}\,,\,F(\scrr, \tau')\,\right]
\,,\qquad\quad\mbox{for $\tau'\;\leq\; \tau$.}
\nonumber\\[1em]
&&\mbox{with adiabatic turn--on initial condition}\quad\lim_{\tau'\to -\infty} W(\scrr\,\vert\,\scrr'(\tau'), \tau') \;=\; 0\,.
\label{eqn-gil-wake-kepW2}
\end{eqnarray}
We now turn to eqn.(\ref{bbgky-gil-kepF}) for the Ring DF $F(\scrr, \tau)$. Since
\beq
\left[\,F(\scrr, \tau)\,,\,\varphi(\scrd, \tau)\,\right]_{(6)} \;=\; - \frac{\p F}{\p I}\frac{\p \varphi}{\p w} \;+\; \left[\,F(\scrr, \tau)\,,\,\varphi(\scrd, \tau)\,\right]\nonumber\,,
\eeq
we have
\beq
\oint\left[\,F(\scrr, \tau)\,,\,\varphi(\scrd, \tau)\,\right]_{(6)}
\,\frac{\rmd w}{2\pi} \;=\; \oint\left[\,F(\scrr, \tau)\,,\,\varphi(\scrd, \tau)\,\right]\,\frac{\rmd w}{2\pi} \;=\; 
\left[\,F(\scrr, \tau)\,,\,\oint\varphi(\scrd, \tau)\frac{\rmd w}{2\pi}\,\right]
\nonumber\,.
\eeq
Also, 
\beq
\left[\,\oint\!\!\oint p(\scrd, \scrd')\frac{\rmd w}{2\pi}\frac{\rmd w'}{2\pi}\,,\, F^{(2)}_{\rm irr}(\scrr, \scrr', \tau)\,\right]_{(6)} \;=\;
\left[\,\oint\!\!\oint p(\scrd, \scrd')\frac{\rmd w}{2\pi}\frac{\rmd w'}{2\pi}\,,\, F^{(2)}_{\rm irr}(\scrr, \scrr', \tau)\,\right]\,.
\nonumber
\eeq
Then eqn.(\ref{bbgky-gil-kepF}) for $F(\scrr, \tau)$
can be written as:
\beq
\frac{\p F}{\p \tau} \;+\; 
\left(1 - \frac{1}{N}\right)\left[\,F\,,\,\oint\varphi(\scrd, \tau)\frac{\rmd w}{2\pi}\,\right] \;=\; 
\frac{1}{N}\int\left[\,\oint\!\!\oint p(\scrd, \scrd')\frac{\rmd w}{2\pi}\frac{\rmd w'}{2\pi}\,,\, F^{(2)}_{\rm irr}(\scrr, \scrr', \tau)\,\right]\,\rmd\scrr'\,.
\label{bbgky-gil-kepF2}
\eeq
Similarly, the solution eqn.(\ref{eqn-gke-kepfn}) for the $O(\varepsilon)$
correction to the DF can be simplified to:
\begin{eqnarray}
f_n(\scrr, \tau) &\;=\;& \frac{1}{\Omega_{\rm k}} \frac{\p F}{\p I}
\oint \varphi(\scrd, \tau)\exp{[-{\rm i}n w]}\,\frac{\rmd w}{2\pi}
\nonumber\\[1ex]
&&\qquad \;+\;
\frac{{\rm i}}{n \Omega_{\rm k}} \left[\,F\,,\,\oint\varphi(\scrd, \tau)\exp{[-{\rm i}n w]}\,\frac{\rmd w}{2\pi}\right]\,,\qquad\qquad n\neq 0\,.
\label{eqn-gke-kepfn2}
\end{eqnarray}

We have now eliminated all the 6--dim PBs and written eqn.(\ref{eqn-gil-wake-kepW2}) for $W$, eqn.(\ref{bbgky-gil-kepF2}) for $F$ and eqn.(\ref{eqn-gke-kepfn2}) for the $f_n$ in terms of 4--dim PBs operating in $\scrr$--space. Only the integrals of various potential functions over $w$ and $w'$ remain to be simplified. Using eqns.(\ref{mean-pot-kep}) for $\varphi\,$,  (\ref{wake-pot-kep}) for $\varphi^{\rm w}$ and $\varphi^{\rm p} = p - \varphi\,$, the integrals 
needed are: 
\begin{eqnarray}
&&\Phi(\scrr, \tau) \;\stackrel{\rm def}{=}\; \oint\varphi(\scrd, \tau)\frac{\rmd w}{2\pi} \;=\; \int F(\scrr', \tau)\,\rmd\scrr'\oint\!\!\oint
p(\scrd, \scrd')\,\frac{\rmd w}{2\pi}\frac{\rmd w'}{2\pi}\,,
\nonumber\\[1em]
&&\varphi_n(\scrr, \tau) \stackrel{\rm def}{=} \oint\varphi(\scrd, \tau)\exp{[-{\rm i}n w]}\,\frac{\rmd w}{2\pi} 
= \int F(\scrr', \tau)\,\rmd\scrr'\oint\!\!\oint
p(\scrd, \scrd')\exp{[-{\rm i}n w]}\,\frac{\rmd w}{2\pi}\frac{\rmd w'}{2\pi}\,,\nonumber\\[1em]
&&\Phi^{\rm w}(\scrr, \scrr', \tau) \;\stackrel{\rm def}{=}\; \oint\varphi^{\rm w}(\scrd, \scrr', \tau)\frac{\rmd w}{2\pi} \;=\; \int W(\scrr''\,\vert\,\scrr', \tau)\,\rmd\scrr''\oint\!\!\oint p(\scrd, \scrd'')\,\frac{\rmd w}{2\pi}\frac{\rmd w''}{2\pi}\,,
\nonumber\\[1em]
&&\Phi^{\rm w}(\scrr, \scrr', \tau) \;\stackrel{\rm def}{=}\; \oint\!\!\oint\varphi^{\rm p}(\scrd, \scrd',\tau')\frac{\rmd w}{2\pi}\frac{\rmd w'}{2\pi} \;=\; \oint\!\!\oint p(\scrd, \scrd'')\,\frac{\rmd w}{2\pi}\frac{\rmd w''}{2\pi} \;-\; \oint\varphi(\scrd, \tau)\frac{\rmd w}{2\pi}\,. 
\nonumber
\end{eqnarray}
A quantity that is common to all these integrals is the partially orbit--averaged Poisson kernel, whose Fourier series we write as: 
\begin{subequations} 
\begin{eqnarray}
\oint p(\scrd, \scrd')\,\frac{\rmd w'}{2\pi} &\;=\;& \Psi(\scrr, \scrr') \;+\; \sum_{n \neq 0}
\psi_n(\scrr, \scrr')\exp{[{\rm i}n w]}\,,
\label{p-exp}\\[1ex]
\mbox{where}\qquad
\Psi(\scrr, \scrr') &\;=\;& -GM_\bullet\oint\oint\frac{{\rm d}w}{2\pi}\,
\frac{{\rm d}w'}{2\pi}\,\frac{1}{\left|\bfr - \bfr'\right|}\,,
\label{Psidef}\\[1ex]
\psi_n(\scrr, \scrr') &\;=\;& -GM_\bullet\oint\oint\frac{{\rm d}w}{2\pi}\,
\frac{{\rm d}w'}{2\pi}\,\frac{\exp{[-{\rm i}n w]}}{\left|\bfr - \bfr'\right|}\,,\qquad\qquad n\neq 0\,,
\label{psindef}
\end{eqnarray}
\end{subequations}
are the Ring--Ring interaction potential functions introduced in Paper~I.
The functions $\Psi$ and $\psi_n$ will be treated as known functions of 
their arguments. Using eqns.(\ref{p-exp})--(\ref{psindef}), we get
\begin{eqnarray}
\Phi(\scrr, \tau) &\;=\;& \int F(\scrr', \tau)\,\Psi(\scrr, \scrr')
{\rm d}\scrr'\,,\label{phislow}\\[1ex]
\varphi_n(\scrr, \tau) &\;=\;& \int F(\scrr', \tau)\,\psi_n(\scrr, \scrr')
{\rm d}\scrr'\,,\label{phin}\\[1ex]
\Phi^{\rm w}(\scrr, \scrr', \tau) &\;=\;& \int \,W(\scrr''\,\vert\,\scrr', \tau)\,\Psi(\scrr, \scrr'')\,{\rm d}\scrr''\,,
\label{phislow-w}\\[1ex]
\Phi^{\rm p}(\scrr, \scrr', \tau) &\;=\;& \Psi(\scrr, \scrr') \;-\; \Phi(\scrr, \tau)\,.
\label{phislow-p} 
\end{eqnarray}

We are now ready to cast eqn.(\ref{eqn-gil-wake-kepW2}) for $W$, eqns.(\ref{bbgky-gil-kepF2}) for $F$ and eqn.(\ref{eqn-gke-kepfn2}) in a form that is determined entirely by $\scrr$--space quantities. The equation for the wake function $W(\scrr\,\vert\,\scrr'(\tau'), \tau')$ is:
\begin{eqnarray}
&&\frac{\p W}{\p \tau'} \;+\; \left[\,W(\scrr\,\vert\,\scrr'(\tau'), \tau')\,,\,\Phi(\scrr, \tau')\,\right]
\;+\; \left[\,F(\scrr, \tau')\,,\,\Phi^{\rm w}(\scrr, \scrr'(\tau'), \tau')\,\right]
\nonumber\\[1em]
&&\qquad\qquad \;=\; \left[\,\Phi^{\rm p}(\scrr, \scrr'(\tau'),\tau')\,,\,F(\scrr, \tau')\,\right]
\,,\qquad\quad\mbox{for $\tau'\;\leq\; \tau$.}
\nonumber\\[1em]
&&\mbox{with adiabatic turn--on initial condition}\quad\lim_{\tau'\to -\infty} W(\scrr\,\vert\,\scrr'(\tau'), \tau') \;=\; 0\,.
\label{eqn-gil-wake-kepW3}
\end{eqnarray}
The kinetic equation for the Ring DF $F(\scrr, \tau)$ is:
\beq
\frac{\p F}{\p \tau} \;+\; 
\left(1 - \frac{1}{N}\right)\left[\,F\,,\,\Phi(\scrr, \tau)\,\right] \;=\; 
\frac{1}{N}\int\left[\,\Psi(\scrr, \scrr')\,,\, F^{(2)}_{\rm irr}(\scrr, \scrr', \tau)\,\right]\,\rmd\scrr'\,.
\label{bbgky-gil-kepF3}
\eeq
The Fourier coefficients of the $O(\varepsilon)$ fluctuations in the DF 
$f_n$ are given by:
\beq
f_n(\scrr, \tau) \;=\; \frac{1}{\Omega_{\rm k}} \left\{\,\frac{\p F}{\p I}
\varphi_n \;+\; \frac{{\rm i}}{n} \left[\,F\,,\,\varphi_n\right]\,\right\}\,,\qquad\qquad n\neq 0\,.
\label{eqn-gke-kepfn3}
\eeq
As earlier the function $f_0(\scrr, \tau)$ remains undetermined by our 
$O(\varepsilon)$ theory. Using eqn.(\ref{eqn-gke-kepfn3}) for 
the $f_n$ in eqn.(\ref{dffou}), we get the following expression for the 
full DF to first order in $\varepsilon$:
\beq
f(\scrd, \tau) \;=\; \frac{1}{2\pi}F(\scrr, \tau) \;+\;
\frac{\varepsilon}{2\pi\Omega_{\rm k}}\,\sum_{n\neq 0}\left(\frac{\p F}{\p I}\varphi_n \;+\; \frac{{\rm i}}{n}
\left[\,F\,,\,\varphi_n\,\right]\right)\exp{[{\rm i}n w]}
\;+\; \frac{\varepsilon}{2\pi} f_0(\scrr, \tau)\,. 
\label{fullsoln}
\eeq
Once the Ring DF $F(\scrr, \tau)$ has been determined by solving 
eqns.(\ref{eqn-gil-wake-kepW4}) and (\ref{bbgky-gil-kepF4}), the 
$O(\varepsilon)$ fluctuations in $w$ are completely determined by 
eqn.(\ref{fullsoln}). To determine the function $f_0(\scrr, \tau)$ it is 
necessary to work to higher order in $\varepsilon$. With this caveat
in mind, the orbit--averaged kinetic description presented here is self--consistent and closed in 5--dim $\scrr$--space.

\end{document}